\documentclass[article]{JHEP3}
\usepackage{amsmath,amssymb}
\usepackage{mathtools}
\usepackage{cite}
\usepackage{verbatim}
\usepackage{float}
\usepackage{subcaption}
\usepackage{amsfonts}
\usepackage[utf8]{inputenc}
\usepackage{graphicx}
\usepackage{nomencl}
\usepackage[a4paper,top=2cm, bottom=2cm, left=2cm, right=2cm]{geometry}

\allowdisplaybreaks

\newcommand{\be}{\begin{equation}}
\newcommand{\ee}{\end{equation}}
\newcommand{\bea}{\begin{eqnarray}}
\newcommand{\eea}{\end{eqnarray}}
\newcommand{\bean}{\begin{eqnarray*}}
\newcommand{\eean}{\end{eqnarray*}}

\def\beq{\begin{equation}}

\def\eeq{\end{equation}}
\def\R{\mathcal{R}}

\def\Re{\mathop{\rm Re}}
\def\Im{\mathop{\rm Im}}
\relax

\def\d{\partial}

\def\a{\alpha'}

\title{Greybody factors of string-corrected $d$-dimensional black holes}

\author{Filipe Moura$^a$ and Jo\~ao Rodrigues$^b$
\\
\\
$^{a}$
Departamento de Matem\'atica, Escola de Tecnologias e Arquitetura, \\ ISCTE - Instituto Universit\'ario de Lisboa \\ and Instituto de Telecomunica\c c\~oes,
\\Av. das For\c cas Armadas, 1649-026 Lisboa, Portugal\\
\email{fmoura@lx.it.pt}
\\
\\
$^{b}$
Centro de An\'alise Matem\'atica, Geometria e Sistemas Din\^amicos,\\ Departamento de Matem\'atica,\\ Instituto Superior T\'ecnico,\\
Av. Rovisco Pais, 1049-001 Lisboa, Portugal\\
\email{joao.carlos.rodrigues@tecnico.ulisboa.pt}
}

\abstract{We compute analytically greybody factors for asymptotically flat spherically symmetric black holes with stringy higher derivative corrections in $d$ dimensions in the high frequency limit. Our calculations include both the eikonal limit - where the real part of the frequency of the scattered wave
is much larger than the imaginary part - and the highly damped case - where the imaginary part of the frequency is much larger than the real part -, addressing the emission of gravitons and test scalar fields, and yielding full transmission and reflection scattering coefficients.
}

\begin{document}



\vfill

\eject

\section{Introduction}
\noindent

In the famous calculation of the Hawking radiation \cite{Hawking:1975vcx}, it was shown that black holes have a thermal spectrum. The expectation value for the number of particles emitted with a certain frequency $\omega$ is
\begin{equation}
    \Big\langle n(\omega) \Big\rangle = \frac{\gamma(\omega)}{e^{\frac{\omega}{T_{\mathcal{H}}}}\pm 1} \label{gfdef}
\end{equation}
where $T_{\mathcal{H}}$ is the Hawking temperature of the black hole space time and the $+/-$ sign addresses radiation composed by fermions or bosons, respectively. The frequency-dependent factor $\gamma(\omega)$ is the greybody factor. Integrating the expression above over the entire frequency spectrum yields the black hole emission rate.

The greybody factor is directly connected to the asymptotic observation of Hawking radiation. Indeed, at the vicinity of the black hole horizon, Hawking radiation is black body radiation. However, as the radiation propagates outside the region bounded by the event horizon it encounters the gravitational potential generated by the black hole itself and scatters on its nontrivial spacetime curvature. This results in the reflection and transmission of the Hawking radiation. The actual spectrum observed by an asymptotic observer is different from a blackbody spectrum. Black hole greybody factors are functions characterizing the tunnelling probability of perturbations through the black hole effective potential \cite{Parikh:1999mf}. A greybody factor is a transmission probability: a quantity that describes the deviation of the Hawking radiation from a pure blackbody radiation. The computation of these factors is therefore crucial in order to understand the Hawking radiation from a semiclassical point of view.

Recently it was observed that black hole greybody factors could also be important in the modelling of post-merger gravitational wave ringdown signals \cite{Oshita:2022pkc,Oshita:2023cjz,Okabayashi:2024qbz,Konoplya:2024lir}. The ringdown phase, describing the relaxation of the remnant black hole, is typically modeled using quasinormal modes, which correspond to exponentially damped oscillations. Although black hole spectroscopy is based on quasinormal modes, these are known to be highly sensitive to relatively small deformations of the black hole geometry. Differently than the quasinormal spectra, greybody factors are stable under such small deformations \cite{Rosato:2024arw,Oshita:2024fzf}. Therefore, greybody factors may be useful not only for computing the spectrum of Hawking radiation but also in the context of astrophysical observations of gravitational waves from black holes.

Greybody factors are directly related to the propagation of fields $\psi$ in a black hole space time. The propagating fields can be any field coupled to gravity, or even linear perturbations of the metric tensor field itself. The simplest case corresponds to a minimally coupled massless scalar field, for which emission spectra have been computed for scalar fields in nonrotating \cite{Page:1976df} and rotating asymptotically flat black holes \cite{Page:1976ki,Starobinsky:1973aij} in $d=4$ Einstein gravity. More recently, these studies have been extended to asymptotically de Sitter black holes, either static \cite{Crispino:2013pya} or rotating \cite{CarneirodaCunha:2015qln} and for near-BPS black holes in $d=4, 5$ \cite{Maldacena:1997ih}. Generic fields have been considered in \cite{Boonserm:2008zg}.

In higher dimensions, the emission spectra has been studied in brane world scenarios, considering the emitted fields as restricted to live on a 4-dimensional brane. With these assumptions, greybody factors have been computed for asymptotically flat \cite{Kanti:2002nr,Harris:2003eg} and de Sitter \cite{Kanti:2014dxa} black holes. The recoil effect on a black hole attached to a brane in a higher dimensional space resulting from the emission of gravitons into the bulk has been studied in \cite{Frolov:2002as} and (considering the effects of brane tension) in \cite{Dai:2006hf}. But greybody factors have also been computed without these assumptions, directly in $d$ dimensions, for asymptotically flat, de Sitter and anti-de Sitter black holes, either static \cite{Harmark:2007jy} or rotating \cite{Jorge:2014kra}. The specific case of graviton emission has been studied in \cite{Cardoso:2005mh}.

Most of the times, greybody factors have to be computed numerically. Nonetheless, different analytical methods have been developed in order to compute them in some limiting cases. One of these limits that is typically considered is the low frequency limit of the emitted fields, associated just to $S$ waves (modes with $\ell=0$, $\ell$ being the multipole number). This limit is also often considered in the related scattering problems by black holes.

A more recent approach to the computation of greybody factors is based on the fact that the dynamics of perturbed non-rotating black holes admits
an infinite number of symmetries that are generated by the flow of the Korteweg-de Vries equation. Associated to these symmetries there is an infinite number of conserved quantities (Korteweg-de Vries integrals), which fully determine the greybody factors. This approach has been applied to Schwarzschild black holes in \cite{Lenzi:2022wjv,Lenzi:2023inn}.


It is of obviously relevance to extend the studies of emission spectra and greybody factors to black holes with higher derivative corrections, namely those coming from string theory. In Einstein-Gauss-Bonnet gravity in $d$ dimensions, greybody factors have been computed, in the low frequency limit and for $S$ waves of scalar, fermion and gauge fields, in brane world scenarios and for asymptotically flat black holes in \cite{Grain:2005my}. This limit was also taken in the study of scattering problems for the same type of black holes with string corrections in \cite{Moura:2006pz,Moura:2011rr}. These studies have been extended for scalar fields and de Sitter $d-$dimensional black holes and for modes with higher $\ell$, also with Gauss-Bonnet corrections, in \cite{Zhang:2017yfu,Zhang:2020qam}. Greybody factors have also been studied for black holes in $f\left(\R\right)$ gravity coupled to a cloud of strings (in $d=3$ \cite{Ovgun:2018gwt}) and for charged black holes in nonlinear electrodynamics (in $d=3$ \cite{Panotopoulos:2018pvu} and $d=4$ \cite{Okyay:2021nnh}).

In this article, we will take the opposite (high frequency) limit in two ways: the eikonal (geometrical optics) limit, corresponding to a large real part of the frequency of the emitted radiation; and the asymptotic (highly damped) limit, corresponding to a large imaginary part of such frequency. We will consider $d-$dimensional spherically symmetric black holes with leading string-theoretical $\a$ corrections, and we will compute their greybody factors for emitted massless scalars and gravitons (corresponding to tensorial perturbations of the metric).

\section{String-corrected spherically symmetric black holes and their gravitational perturbations}
\label{sstp}
\noindent

A general static spherically symmetric metric in $d$ dimensions can always be cast in the form
\be \label{schwarz}
ds^2 = -f(r)\ dt^2  + f^{-1}(r)\ dr^2 + r^2 d\Omega^2_{d-2}.
\ee
The tortoise coordinate $x$ for the metric (\ref{schwarz}) is defined by
\be
dx = \frac{dr}{f(r)}. \label{tort}
\ee

In the background of a spacetime of the form (\ref{schwarz}), any scalar field $\Psi(t,r,\theta)$ can be expanded as
\begin{equation}
\Psi(t,r,\theta)= e^{i\omega t} \sum_{\ell} \psi_{\ell, \omega}(r) Y_{\ell}(\theta)\,,
\label{sphericalharmonics1}
\end{equation}
where $\omega$ is the wave frequency, $\ell$ is the angular quantum number associated with the polar angle $\theta$ and $Y_{\ell} (\theta)$ are the usual spherical harmonics defined over the $(d-2)$ unit sphere $\mathbb{S}^{d-2}$. If $\Psi(t,r,\theta)$ given by (\ref{sphericalharmonics1}) is a minimally coupled massless test scalar field, the Klein-Gordon equation $\nabla^\mu \nabla_\mu \Psi=0$ becomes separable in angular and radial parts, each one obeying its second order differential equation. Concretely for the radial part, each component $\psi_{\ell, \omega}(r)$ (for simplicity, $\psi(r)$) obeys the field equation
\be
\frac{d^2 \psi}{d\, x^2} + \omega^2 \psi = V \left[ f(r) \right] \psi \label{potential0}
\ee
with a potential given by \cite{Harmark:2007jy}
\be
V_{\textsf{min}} [f(r)] = f(r) \left( \frac{\ell \left( \ell + d - 3 \right)}{r^2} + \frac{\left( d - 2 \right) \left( d - 4 \right) f(r)}{4r^2} + \frac{\left( d - 2 \right) f'(r)}{2r} \right). \label{v0}
\ee

General tensors of rank at least 2 on $\mathbb{S}^{d-2}$ can be uniquely decomposed in their tensorial, vectorial and scalar components. That is the case, for instance, of general perturbations $h_{\mu\nu}=\delta g_{\mu\nu}$ of a $d-$dimensional spherically symmetric metric like (\ref{schwarz}: we have then scalar, vectorial and (for $d>4$) tensorial gravitational perturbations. Each type of perturbation is described in terms of master variables, which we also designate generically by $\psi(r)$.
Each of these master variables obeys \cite{ik03a} a second order differential equation (``master equation'') with a potential, like (\ref{potential0}). The potential $V \left[ f(r) \right]$ in this case depends on the kind of perturbation, and also on the Lagrangian one considers. In Einstein gravity, for tensorial perturbations of the metric in (\ref{potential0}), the potential is given by $V_{\textsf{min}}$ in (\ref{v0}) \cite{ik03a}: it coincides with the potential for test scalar fields.

In the presence of higher order corrections in the Lagrangian, one can still have spherically symmetric black holes of the form (\ref{schwarz}), but the master equation obeyed by each perturbation variable is expected to change. Concretely, we will consider the following $d$--dimensional effective action with leading string-theoretical $\a$ corrections:
\be \label{eef} \frac{1}{16 \pi G} \int \sqrt{-g} \left( \R -
\frac{4}{d-2} \left( \d^\mu \phi \right) \d_\mu \phi +
\mbox{e}^{\frac{4}{d-2} \phi} \frac{\lambda}{2}\
\R^{\mu\nu\rho\sigma} \R_{\mu\nu\rho\sigma} \right) \mbox{d}^dx .
\ee
This is the effective action of bosonic and heterotic string theories, to first order in the inverse string tension $\a$, with $\lambda = \frac{\a}{2}, \frac{\a}{4}$, respectively. \footnote{Type II superstring theories do not have $\a$ corrections to this order.} In both cases, since we are only interested in purely gravitational corrections, we can consistently set all other bosonic and fermionic fields present in the string spectrum to zero except for the dilaton field $\phi$.

In \cite{Moura:2006pz,Moura:2012fq} it has been shown that, perturbing the field equations resulting from this action, for tensorial perturbations of the metric (\ref{schwarz}) one also obtains a second order master equation like (\ref{potential0}). The corresponding potential, as expected, is an $\a$-corrected version of the minimal potential in (\ref{v0}) given by
\bea
V_{\textsf{T}} [f(r)] &=& \lambda\ \frac{f(r)}{r^2} \left[ \left( \frac{2 \ell \left( \ell + d - 3 \right)}{r} + \frac{\left( d - 4 \right) \left( d - 5 \right) f(r)}{r} + \left( d - 4 \right) f'(r) \right) \left( 2 \frac{1 - f(r)}{r} + f'(r) \right) \right. \nonumber \\
&+& \left. \Big( 4 (d-3) - (5d-16) f(r) \Big) \frac{f'(r)}{r} - 4 \left( f'(r) \right)^2 + \left( d-4 \right) f(r) f''(r) \right] + V_{\textsf{min}} [f(r)]. \label{vt}
\eea

Spherically symmetric $d-$dimensional black hole solutions with these $\a$ corrections have been obtained in \cite{cmp89,Moura:2009it}. Specifically concerning the action (\ref{eef}), a solution of the respective field equations is of the form (\ref{schwarz}), with

\bea \label{fr2}
f(r) &=& f_0(r) \left(1+ \frac{\lambda}{R_H^2} \delta f(r) \right), \\
f_0(r) &=& 1 - \frac{R_H^{d-3}}{r^{d-3}}, \label{fr0}\\
\delta f(r) &=& - \frac{(d-3)(d-4)}{2}\ \frac{R^{d-3}_H}{r^{d-3}}\ \frac{1 - \frac{R_H^{d-1}}{r^{d-1}}}{1 - \frac{R^{d-3}_H}{r^{d-3}}}.
\eea
The only horizon of this metric occurs at the same radius $r=R_H$ of the Tangherlini solution, which is the metric with $f(r)=f_0(r)$ obtained in the Einstein limit $\lambda=0$.

This asymptotically flat black hole solution has been obtained by Callan, Myers and Perry in \cite{cmp89}, where some of its properties have been studied. For our purposes, it is enough to quote here the explicit expression for its temperature, given by
\be
T_{\mathcal{H}} = \frac{f'(R_H)}{4 \pi}= \frac{f_0'(R_H)}{4 \pi} \left(1+ \frac{\lambda}{R_H^2} \delta f(R_H) \right)= \frac{d-3}{4 \pi R_H} \left( 1 - \frac{\left( d-1 \right) \left( d-4 \right)}{2}\ \frac{\lambda}{R_H^2} \right). \label{temp}
\ee

Throughout this article we will use for the perturbative expansion the small dimensionless parameter
\be
\lambda' = \frac{\lambda}{R_H^2}= \lambda \frac{16 \pi^2}{(d-3)^2} T_\mathcal{H}^2. \label{lprime}
\ee

\begin{figure}[h]
\begin{subfigure}{0.32\textwidth}
\includegraphics[width=5.2cm, height=5.2cm]{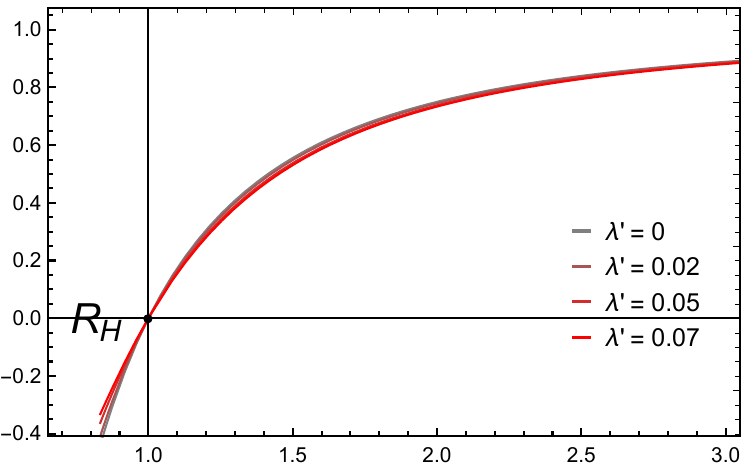}
\caption{$d=5$}
\end{subfigure}
\begin{subfigure}{0.32\textwidth}
\includegraphics[width=5.2cm, height=5.2cm]{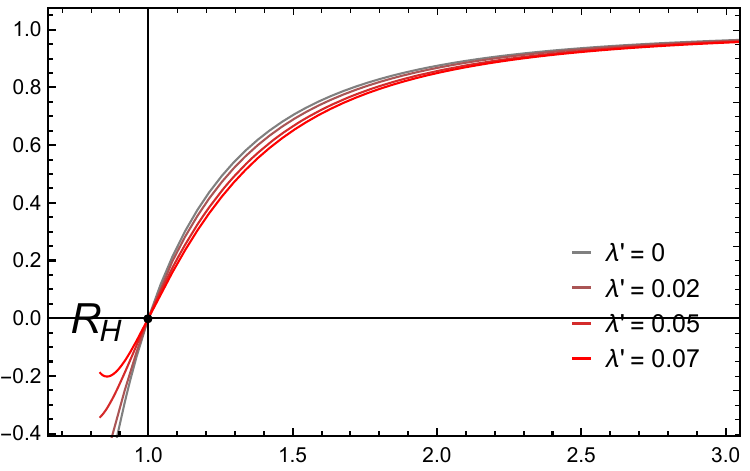}
\caption{$d=6$}
\end{subfigure}
\begin{subfigure}{0.32\textwidth}
\includegraphics[width=5.2cm, height=5.2cm]{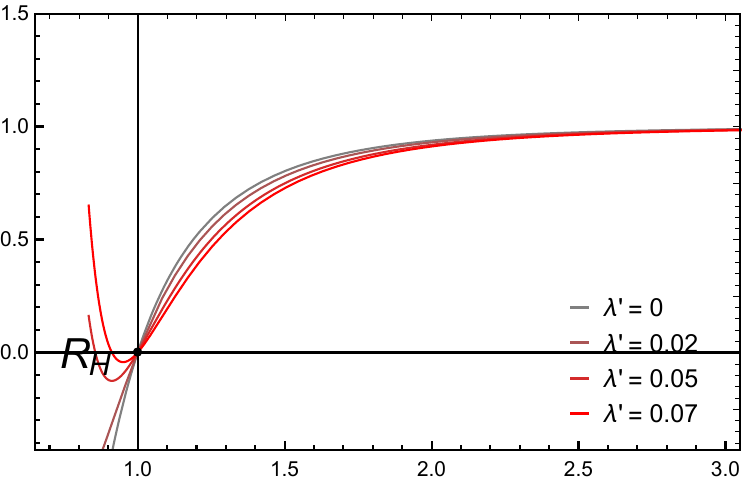}
\caption{$d=7$}
\end{subfigure}
\caption{Plot of the metric function $f(r)$ for different dimensions and values of $\lambda'$. The horizontal axis stands for $r/R_H$.}
\label{fig1}
\end{figure}

\begin{figure}[h]
\begin{subfigure}{0.32\textwidth}
\includegraphics[width=5.2cm, height=5.2cm]{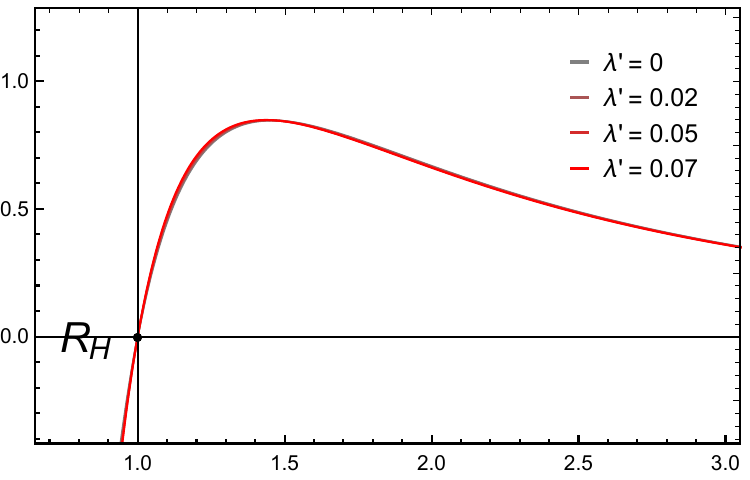}
\caption{$d=5$}
\end{subfigure}
\begin{subfigure}{0.32\textwidth}
\includegraphics[width=5.2cm, height=5.2cm]{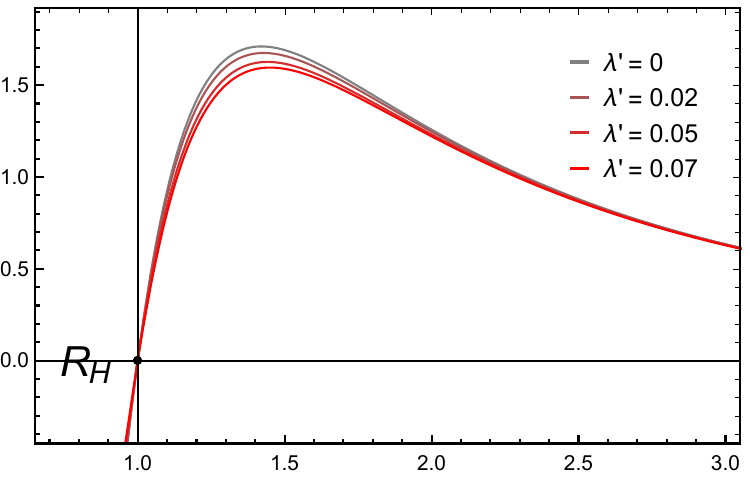}
\caption{$d=6$}
\end{subfigure}
\begin{subfigure}{0.32\textwidth}
\includegraphics[width=5.2cm, height=5.2cm]{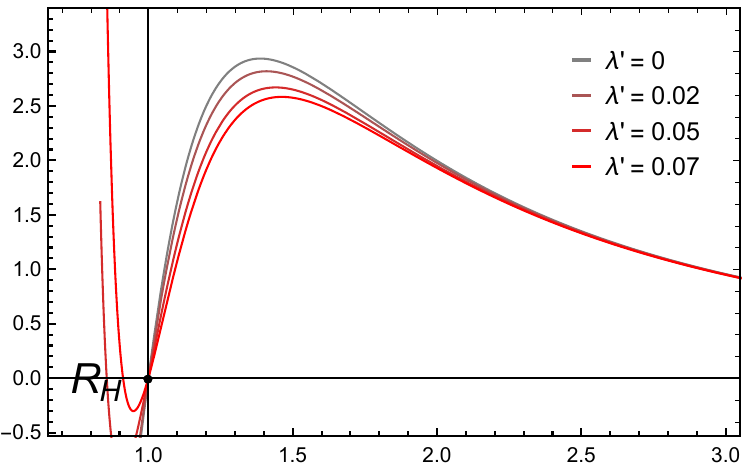}
\caption{$d=7$}
\end{subfigure}
\caption{Plot of the potential $V_{\textsf{min}} [f(r)]$ for different dimensions and values of $\lambda'$. The horizontal axis stands for $r/R_H$.}
\label{fig5}
\end{figure}

\begin{figure}[h]
\begin{subfigure}{0.32\textwidth}
\includegraphics[width=5.2cm, height=5.2cm]{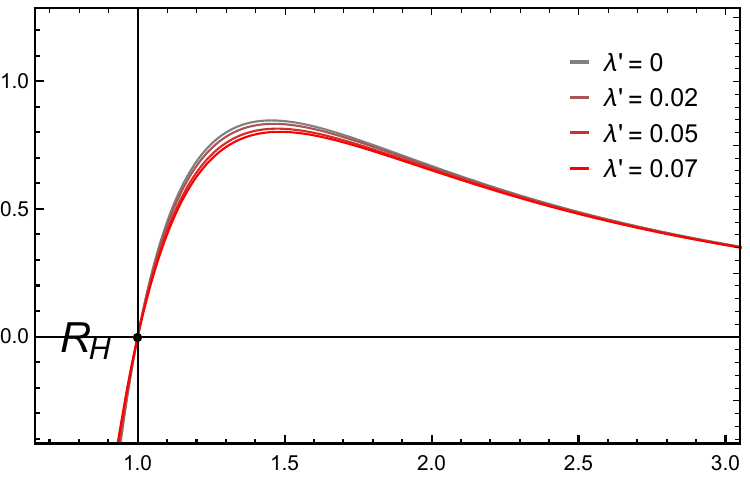}
\caption{$d=5$}
\end{subfigure}
\begin{subfigure}{0.32\textwidth}
\includegraphics[width=5.2cm, height=5.2cm]{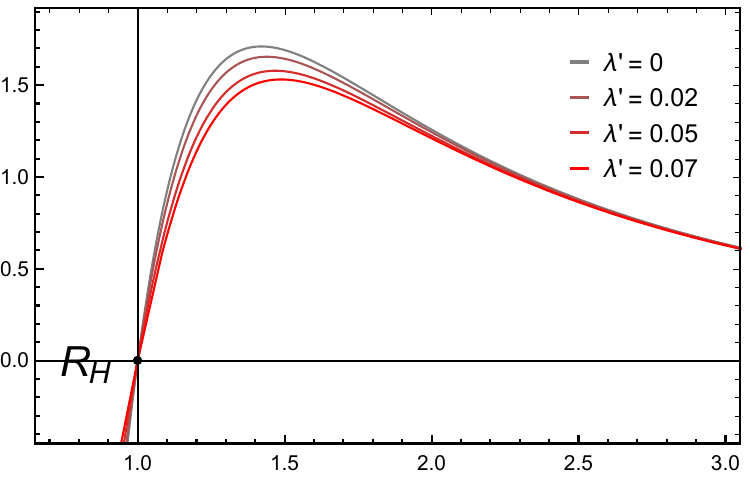}
\caption{$d=6$}
\end{subfigure}
\begin{subfigure}{0.32\textwidth}
\includegraphics[width=5.2cm, height=5.2cm]{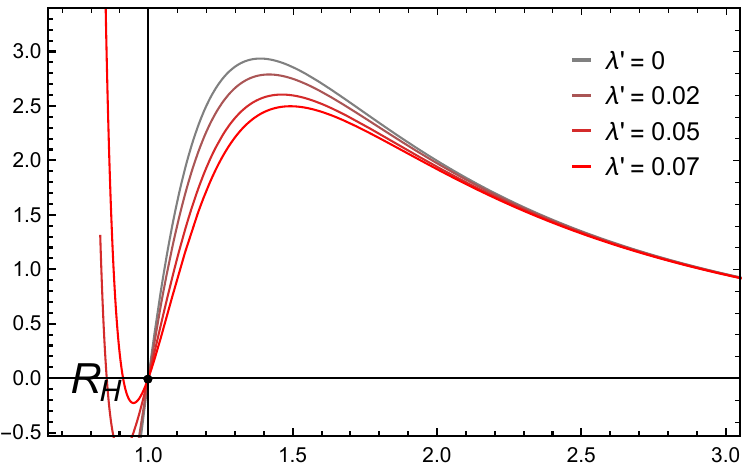}
\caption{$d=7$}
\end{subfigure}
\caption{Plot of the potential $V_{\textsf{T}} [f(r)]$ for different dimensions and values of $\lambda'$. The horizontal axis stands for $r/R_H$.}
\label{fig6}
\end{figure}

A plot of the metric function $f(r)$ given by (\ref{fr2}) in $d=5, \,6, \,7$ and in units where $R_H=1$ can be seen in figure \ref{fig1} for $\lambda'=0, \lambda'=0.02, \lambda'=0.05, \lambda'=0.07.$ Using this same metric function $f(r)$ and for the same values of $d$ and $\lambda'$ we also present in figures \ref{fig5} and  \ref{fig6} plots of the potentials $V_{\textsf{min}} [f(r)]$ for minimally coupled massless scalar fields and $V_{\textsf{T}} [f(r)]$ for tensorial gravitational perturbations. In all cases we see that the introduction of the string-theoretical perturbative parameter $\lambda'$ does not lead to substantial qualitative changes in the shape of these functions, although the quantitative changes seem to become more significant with increasing $d$. But the presence of $\lambda'$ (or $\a$) leads to changes in the values of physical quantities appearing as perturbative corrections.

The stability of this black hole solution under tensorial gravitational perturbations has been studied in \cite{Moura:2006pz}, and the spectra of quasinormal modes corresponding to such perturbations and also of test scalar fields has been obtained in \cite{Moura:2021eln}, in the eikonal limit, and in \cite{Moura:2021nuh}, in the highly damped regime.

\section{Greybody factors}
\noindent

The tortoise coordinate $x$ is defined in (\ref{tort}) in such a way that, for asymptotically flat black holes, $r\to + \infty$ corresponds to $x\to + \infty$ and $r\to R_H$ corresponds to $x\to - \infty$. In these regions, the potential $V \left[ f(r) \right]$ in (\ref{potential0}) should go to zero:
\bea
    \lim_{r\to + \infty}V \left[ f(r) \right] = 0, \\
    \lim_{r\to + R_H}V \left[ f(r) \right] = 0.
\eea
That is the case of the potentials $V_{\textsf{min}}$ in (\ref{v0}) and $V_{\textsf{T}}$ in (\ref{vt}) we have mentioned, and it should be of other potentials created by the black hole that we could consider (namely corresponding to the other kinds of gravitational perturbations). With these conditions, asymptotically in the regions $x\to - \infty$ and $x \to+ \infty$ $\psi(x)$ should have an oscillatory behavior:
\begin{equation}
    \psi(x) \sim A_+ e^{i\omega x} + A_-e^{-i\omega x} \label{expterms}
\end{equation}
for some constants $A_\pm$. Having this in mind, in order to obtain the greybody factor we seek solutions $\psi_\omega(x)$ of (\ref{potential0}) with frequency $\omega$ that obey the boundary conditions
\bea
\psi_\omega(x) \sim T(\omega) e^{i\omega x}, \, x\to - \infty, \label{129} \\
\psi_\omega(x) \sim e^{i\omega x}+ R(\omega)e^{-i\omega x}, \, x\to + \infty, \label{130}
\eea
where the complex numbers $T(\omega), R(\omega)$ are called transmission and reflection coefficients respectively.

Physically, $\psi_\omega(x)$ describes the scattering of an incoming wave originating at spatial infinity. In general, a scattering problem with a potential allows for solutions with complex frequency $\omega$. In this case, the real part of $\omega$ represents the proper frequency of the wave, and the imaginary part represents its damping. From the decomposition (\ref{sphericalharmonics1}) we see that in order to have damped solutions one should have $\Im\left(\omega\right)>0.$ We will always assume that requirement - otherwise, we would have an unstable solution (and an instability of the black hole, in the case of the scattered wave representing a graviton associated with a perturbation of the black hole).

Because the frequency $\omega$ is complex, we should carefully consider the solution $\psi_{-\omega}(x)$ (see \cite{Harmark:2007jy} for details). We see that $\psi_{-\omega}$ satisfies exactly the same equation (\ref{potential0}) as $\psi_\omega$, since this equation is invariant under the transformation $\omega \mapsto -\omega$, but with boundary conditions
\begin{eqnarray}
\psi_{-\omega} &\sim& e^{-i \omega x} + \widetilde{R} e^{i\omega x}, \qquad x \to + \infty, \label{129til} \\
\psi_{-\omega} &\sim& \widetilde{T} e^{-i \omega x}, \qquad x \to - \infty, \label{130til}
\end{eqnarray}
for some other reflection and transmission coefficients $\widetilde{R}(\omega) := R(-\omega)$, $\widetilde{T}(\omega) := T(-\omega).$ One can easily show that the flux (or wronskian) $J = \frac{1}{2i} \left( \psi_{-\omega}\frac{d \psi_\omega}{dx} - \psi_{\omega}\frac{d \psi_{-\omega}}{dx} \right)$ does not depend on $x$, i.e. $\frac{d J}{dx}=0$. Evaluating such flux at both $x \to \pm \infty$ yields the condition (valid for all asymptotically flat space times \cite{Harmark:2007jy})
\begin{equation}
R \widetilde{R} + T \widetilde{T} = 1. \label{148}
\end{equation}
The greybody factor is defined as
\begin{equation}
    \gamma(\omega) = T(\omega)\widetilde{T}(\omega).
    \label{149}
\end{equation}

If the frequency $\omega$ is real, $\psi_{-\omega}=\psi_{\omega}^*$, $\widetilde{R}= R^\ast$ and $\widetilde{T} =T^\ast$. One has in this case the familiar formulas
\begin{equation}
\left|R(\omega)\right|^2 + \left|T(\omega)\right|^2 = 1, \, \,
\gamma(\omega) = \left|T(\omega)\right|^2.
\label{gfd}
\end{equation}

Physically, the computation of a greybody factor models the scattering of a propagating field in the non trivial structure of the potential $V$, resultant from the curvature of space time. Moreover, despite we consider the emitted radiation with complex frequency, this scattering problem itself differs from the calculation of quasinormal modes in two main ways and should be clearly distinguished from it. First of all, the functional form of the boundary condition at spatial infinity (\ref{130}) is different. Physically, the $e^{i\omega x}$ term in (\ref{130}) means we allow propagating waves arriving from spatial infinity, as opposed to the quasinormal modes problem (where that term is absent in the corresponding boundary condition). Secondly, we care about the constants $T(\omega), R(\omega)$ multiplying the exponential functions on the boundary conditions (\ref{129}) and (\ref{130}). This is so, because they play a role in the physical interpretation and relevance of the problem. In the computation of quasinormal modes this is not the case, for the functional form of the boundary conditions alone is enough to yield plenty of information about the quasinormal frequencies.

When defining the greybody factor $\gamma(\omega)$, only for real frequency $\omega$ of the emitted radiation it makes sense to integrate Hawking's formula (\ref{gfdef}) over the whole frequency spectrum and talk about a radiation emission rate. In general there is some arbitrariness in continuing (\ref{gfdef}) to complex $\omega.$ Since the imaginary part of $\omega$ represents a radiation damping, in the limit of pure imaginary $\omega$ like we take one can think of the integral of formula (\ref{gfdef}) over the whole frequency spectrum as a radiation decay rate.

In this article, we compute analytical expressions for the greybody factors associated with gravitons corresponding to tensor type gravitational perturbations and with test scalar fields in the $d$-dimensional black hole space time with leading string $\a$ corrections obtained by Callan, Myers and Perry in \cite{cmp89}. We consider two different limits: the eikonal limit of large $\ell$ and the asymptotic limit $|\Im(\omega)|\gg 1$.

\section{Greybody factors in the eikonal limit}
\noindent

In the eikonal limit one can use the WKB method in order to solve scattering problems described by equations like (\ref{potential0}), as long as the associated potential $V\left[f(r)\right]$ has one single peak (maximum). That is the case for the potentials corresponding to test scalar fields and tensorial gravitational perturbations in Einstein gravity, and also in the presence of string corrections. Indeed, as $\lambda'$ is a small perturbative parameter, the string $\a$ corrections we consider do not change the shape of these potentials.

Greybody factors in the eikonal limit, for spherically symmetric $d-$dimensional black holes in Einstein gravity, have been obtained in \cite{Konoplya:2019hlu,Konoplya:2023moy}. The method relies on a function $U(x,\omega)=V(x)-\omega^2$ and its derivatives, $V(x)$ being the potential $V\left[f(r)\right]$ in (\ref{potential0}), but given in terms of the tortoise coordinate $x$. If $x_0$ is the point where the potential $V(x)$ reaches a maximum, we define
\be
U_{0}(\omega) = U(x_0,\omega), \, U_2(\omega) = \left.\frac{d^2 \,U(x,\omega)}{d\,x^2}\right|_{x=x_0}. \label{uw}
\ee
The greybody factor is then simply given by
\be
\gamma(\omega)=\frac{1}{1+e^{2\pi i \kappa}}, \,\, \kappa=-i\frac{U_{0}(\omega)}{\sqrt{-2 U_2(\omega)}}. \label{gkk}
\ee
According to our perturbative approach, $\gamma(\omega)$ can be expanded in $\lambda'$ as
\begin{equation}
    \gamma(\omega) = \gamma_0(\omega)\left(1+\lambda' \delta\gamma(\omega)\right). \label{geik}
\end{equation}
In this expression, $\gamma_0$ corresponds to the eikonal greybody factor in Einstein gravity, without any $\lambda'$ corrections; it depends only on the eikonal limit of the uncorrected potential $V_{\textsf{min}} [f(r)]$ given by (\ref{v0}). This limit is the same for all gravitational perturbations and scalar test fields \cite{ik03a}. The result for $\gamma_0$ is therefore universal, and given by
\be
\gamma_0(\omega) = \left(\exp \left(\frac{\pi \ell}{\sqrt{d-3}}-\frac{4^{\frac{1}{3-d}-2} \sqrt{d-3} (d-1)^{\frac{2}{d-3}+1}}{\pi \ell} \frac{\omega^2}{T_H^2} \right)+1\right)^{-1}. \label{geik0}
\ee
In the eikonal (large $\ell$) limit we are working, besides the $\lambda'$ expansion one can also consider an expansion in $1/\ell$. We have then $1/\ell$ corrections to the $\lambda'$ correction $\delta \gamma(\omega)$ in (\ref{geik}), which can be written as
\begin{equation}
    \delta \gamma(\omega) = \Gamma_1(\omega)\left(\delta\Gamma_0(\omega)\ell + \delta\Gamma_1(\omega)\frac{1}{\ell} + \mathcal{O}\left(\frac{1}{\ell^2}\right) \right). \label{dgeik}
\end{equation}
The term $\Gamma_1(\omega)$ is also the same for gravitational perturbations and scalar test fields, and given by
\be
\Gamma_1(\omega) = \left(\exp \left(\frac{4^{\frac{1}{3-d}-2} \sqrt{d-3} (d-1)^{\frac{2}{d-3}+1}}{\pi \ell} \frac{\omega^2}{T_H^2}-\frac{\pi  \ell}{\sqrt{d-3}}\right)+1\right)^{-1}. \label{geik1}
\ee
The corrections $\delta\Gamma_0(\omega), \, \delta\Gamma_1(\omega)$ depend on the potential in (\ref{potential0}). We will compute them separately for the two cases we consider.

\subsection{Tensorial gravitational perturbations}
\noindent

Tensorial gravitational perturbations correspond to the potential (\ref{vt}). The eikonal limit of this potential, and the point where it reaches the maximum value, have been studied in \cite{Moura:2021eln}. The corresponding functions $U_{0}(\omega), U_{2}(\omega)$ defined in (\ref{uw}) are given for this potential by
\bea
U_{0}(\omega) &=& \frac{2^{\frac{2}{d-3}} (d-3) \ell^2}{(d-1)^{\frac{2}{d-3}+1} R_H^2} \left(\left(3 \times 2^{\frac{2}{d-3}+1} (d-2) (d-1)^{-\frac{2}{d-3}-1}-d+4\right) \lambda' +1\right)-\omega^2, \\
U_2(\omega) &=& \frac{2^{\frac{d}{d-3}} (d-3)^3 \ell^2}{(d-1)^{\frac{7}{d-3}+3} R_H^4} \left(2^{\frac{d}{d-3}} (d-7) (d-2) (d-1)^{\frac{1}{d-3}} \lambda' +2^{\frac{1}{d-3}} (d-1)^{\frac{d}{d-3}} (2 (d-4) \lambda' -1)\right).
\eea
From these values we can compute $\gamma(\omega)$ in (\ref{gkk}), namely the missing corrections in (\ref{dgeik}). These are given by
\bea
\delta\Gamma_0(\omega) &=& \frac{\pi}{\sqrt{d-3}} \frac{3\times 4^{\frac{5 d}{d-3}}-4^{\frac{15}{d-3}+4} ((d-3) d+14)}{ 2^{\frac{28}{d-3}+8} (d-1)^{\frac{2}{d-3}+1}}, \\
\delta\Gamma_1(\omega) &=& \frac{\omega^2}{T_H^2} \frac{2^{\frac{3-8 d}{d-3}} (d-1)^{-\frac{2 (d+2)}{d-3}}}{\pi \sqrt{d-3}} \nonumber \\
&& \left[\left(2^{\frac{21}{d-3}+4} \left(((d-14) d+66) d^3+125 d-42\right)-17\times 2^{\frac{7 d}{d-3}} d^2\right) (d-1)^{\frac{10}{d-3}} \right.
\nonumber \\ &&\left. -2^{\frac{19}{d-3}+4} (d-4) (d-3) (d-1)^{\frac{4 d}{d-3}}+2^{\frac{19}{d-3}+4} (d-4) (d-3) (d-1)^{\frac{3 (d+1)}{d-3}} \right].
\eea
\subsection{Test scalar fields}
\noindent

Scalar test fields correspond to the potential (\ref{v0}). The eikonal limit of this potential, and the point where it reaches the maximum value, have also been studied in \cite{Moura:2021eln}. The functions $U_{0}(\omega), U_{2}(\omega)$ in (\ref{uw}) are given for this potential by
\bea
U_0(\omega) &=& \frac{2^{\frac{1}{d-3}} (d-3) \ell^2}{(d-1)^{\frac{4}{d-3}+2} R_H^2} \left(2^{\frac{1}{d-3}} (d-1)^{\frac{2}{d-3}+1} (1-(d-4) \lambda' )+2^{\frac{d}{d-3}} (d-4) \lambda'\right)-\omega ^2, \\
U_2(\omega) &=& \frac{2^{\frac{d}{d-3}} (d-3)^3 \ell^2}{(d-1)^{\frac{6}{d-3}+3} R_H^4} \left(2^{\frac{1}{d-3}} (d-1)^{\frac{2}{d-3}+1} (2 (d-4) \lambda' -1)+2^{\frac{d}{d-3}} (d-4) (d-3) \lambda'\right).
\eea
From these values we can compute the missing corrections to $\gamma(\omega)$ in (\ref{dgeik}). These are given by
\bea
\delta\Gamma_0(\omega) &=&-\frac{\pi 2^{\frac{2}{d-3}} (d-6) (d-1)^{-\frac{2}{d-3}}}{\sqrt{d-3}}, \\
\delta\Gamma_1(\omega) &=& \frac{\omega ^2}{T_H^2} \frac{\sqrt{d-3}}{(d-1)^{\frac{4}{d-3}} 2^{\frac{2}{d-3}+4} \pi} \nonumber \\ &&\left(2^{\frac{2}{d-3}} (d-7) (d-2) (d-1)^{\frac{4}{d-3}}-(d-4) (d-1)^{\frac{2 d}{d-3}}+(d-4) (d-1)^{\frac{d+3}{d-3}}\right).
\eea

From scattering theory one expects, for very high (real) frequencies (corresponding to very high energies), the reflection coefficient $R(\omega)$ to vanish, and hence the greybody factor $\gamma(\omega)$ to be 1. This should be valid already in Einstein gravity, which means that one should have $\lim_{\omega\to + \infty} \gamma_0(\omega)= 1$; therefore, from (\ref{geik}), in this limit the $\lambda'$ corrections should vanish: $\lim_{\omega\to + \infty} \delta \gamma(\omega) = 0$. As one can easily check from (\ref{geik0}), (\ref{geik1}) and the obtained expressions for $\delta\Gamma_0(\omega), \delta\Gamma_1(\omega)$, this is the case for the results we have found in this section, both for test scalar fields and for tensorial gravitational perturbations.

\section{Greybody factors in the asymptotic limit}
\noindent

We now turn to the calculation of the greybody factors in the asymptotic (highly damped) regime, where $|\Im(\omega)|\gg 1$.

\subsection{General setup}
\label{gs}
\noindent

In order to compute the greybody factors in the asymptotic limit, we use the monodromy method. As such, we reuse plenty of definitions and results from our previous article \cite{Moura:2021nuh}, where we computed the asymptotic quasinormal frequencies associated with the same black hole spacetime (\ref{fr2}). The general outline of this procedure will closely follow \cite{Neitzke:2003mz,Keshet:2007be}.

Computing greybody factors requires one to gather information from the boundary conditions, in this case given by (\ref{129}) and (\ref{130}). In the case of complex frequency $\omega$, the boundary conditions (\ref{129}) and (\ref{130}) are very difficult to impose, both for numerical and analytical approximate methods, just as in the quasinormal modes problem. This is because, due to the complex frequency, we are dealing with an exponential large and an exponential small term on the boundaries of the problem. This would result in an indeterminacy in the definition of the asymptotic solutions: the Stokes phenomenon.

This issue can be partially solved in the same way we solved the analogous issue for the computation of quasinormal modes \cite{Moura:2021nuh,Moura:2022gqm}. Indeed, one can allow $r$ to take complex values and consequently assuming the analytic continuation of every function of $r$ to the complex plane. In this case, we can gather information from the boundary condition (\ref{129}) by appealing to the monodromy of $\psi$ around the event horizon.

In order to deal with this problem we should take the contour around a Stokes line in the complex $r$ plane, where the Stokes phenomenon is not manifest. In our case, such lines are defined by the condition $\Im\left(\omega x\right)=0$; since we are considering the asymptotic limit of large imaginary frequencies, this condition can be replaced by $\Re(x)=0$. Through Stokes lines we have then in this case $|e^{\pm i\omega x}| = 1$: the asymptotic behavior of $e^{\pm i\omega x}$ is always oscillatory and there will be no problems with exponentially growing versus exponentially vanishing terms in (\ref{130}). Thus if one considers the Stokes lines, imposing the boundary condition (\ref{130}) in the complex $r$ plane no longer poses a challenge to an approximate analytical method. This, in turn, will allow us to match solutions to the wave equation along the contour $\Im\left(\omega x\right)=0$, even when these were found in very different physical regions.

As we have shown in \cite{Moura:2021nuh}, there is a singularity in the coordinate $x$ at $r=0$ associated to the $\a$ correction.
Because of such singularity, close to the origin the Stokes lines $\Re(x)=0$ are very difficult to handle. Since the analysis of these lines is crucial for our calculation, we must use an alternative coordinate in order to avoid that singular behavior close to the origin. Bacause of that, we rather take the tortoise coordinate $z$ corresponding to the Tangherlini solution, since there are no $\a$ corrections associated to it. Such coordinate is given simply by
\begin{equation}
dz = \frac{dr}{f_0(r)} \label{tort2}
\end{equation}
with $f_0(r)$ given by (\ref{fr0}). Stokes lines associated to this coordinate, given by $\Re(z)=0$, are much easier to handle. A detailed discussion of these Stokes lines, including plots with their complete depiction for every relevant dimension $d$, can be found in \cite{Moura:2021nuh}. In general we will have $2d-4$ Stokes lines emerging from the origin of the complex $r$-plane, all equally distributed and separated by an angle of $\frac{\pi}{d-2}$. Two of such lines are bounded, forming angles of $\pm \frac{\pi}{2(d-2)}$ with the real axis at the origin, and forming a loop around the real physical horizon $r=R_H$. The next two adjacent Stokes lines are unbounded, going towards complex infinity and forming angles of $\pm \frac{3 \pi}{2(d-2)}$ with the real axis at the origin. Between these two unbounded Stokes lines there will be no roots of the metric function $f(r)$ (so called ``fictitious'' horizons).

The general idea of the monodromy method is to pick two closed homotopic contours on the complex $r$-plane. Both these contours enclose only the physical horizon $r=R_H$: none of them encloses the origin of the complex $r$-plane nor any other complex root of the metric function $f(r)$ (``fictitious horizon''). One of these contours, the ``big contour'', seeks to encode information of the boundary condition (\ref{130}) on the monodromy of $\psi$ associated with a full loop around it. The other contour, the ``small contour'', seeks to encode information of the boundary condition (\ref{129}) on the monodromy of $\psi$ associated with a full loop around it. Since both contours are homotopic, the monodromy theorem asserts that the respective monodromies must be the same. Equating them yields an analytic condition which, together with the boundary conditions, allows to compute the greybody factor.

\begin{figure}[h]
\centering
\includegraphics[width=0.5\textwidth]{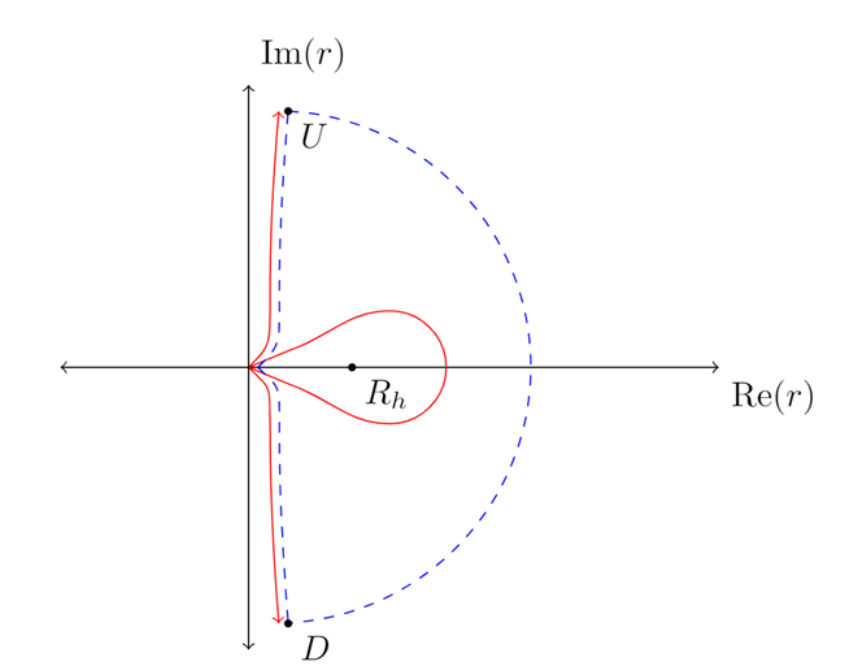}
\caption{Schematic depiction of the big contour, as the blue dashed line. Red curves are Stokes lines. Furthermore, we marked by $D$ and $U$ the regions where the boundary condition (\ref{130}) may be imposed.}
\label{fig2}
\end{figure}

Overall, the big contour is well represented as depicted in figure \ref{fig2} for every dimension $d \geq 5$.

The boundary condition (\ref{130}) is to be imposed in the regions marked by $D$ or $U$. Here, we choose the region $D$ to impose it. Furthermore, we choose to follow the contour in the clockwise direction.

In order to solve the differential equation (\ref{potential0}), we apply standard perturbation theory in $\lambda'$. To start, we must write the differential equation in terms of $z$. From (\ref{tort}), (\ref{fr2}) and (\ref{tort2}), (\ref{potential0}) can be written as
\begin{equation}
\left(\frac{dz}{dx}\right)^2 \frac{d^2 \psi}{dz^2} + \frac{d^2z}{dx^2} \frac{d\psi}{dz} +\left( \omega^2 -V\right)\psi=0,
  \label{potentialz}
\end{equation}
with $\frac{dz}{dx}$ and $\frac{d^2z}{dx^2}$ given as functions of $r$ by
\bea
\frac{dz}{dx} &=& 1+ \frac{\lambda}{R_H^2} \delta f(r); \label{dzdx} \\
\frac{d^2z}{dx^2} &=& f \frac{d}{dr}\left(\frac{dz}{dx} \right) = \frac{\lambda}{R_H^2} f(r) \left(\delta f(r)\right)' \label{d2zdx2}.
\eea
Taking a solution with frequency $\omega$ to (\ref{potential0}) (and also to (\ref{potentialz})) $\psi_\omega$, we also expand it, together with the potential $V$, to first order in $\lambda'$ as
\bea
\psi_\omega &=& \psi_0 + \lambda' \psi_1, \label{psidef}\\
V[f(r)] &=& V_0(r) + \lambda' V_1(r). \label{v1def}
\eea
The $\lambda'=0$ part of the full potential $V[f(r)]$ is given by $V_0(r)=V_{\textsf{min}} [f_0(r)]$, with $V_{\textsf{min}}$ given by (\ref{v0}): it is the classical (uncorrected) potential evaluated with the uncorrected metric function (\ref{fr0}). All the $\lambda'$ corrections appear in $V_1(r)$: those that are implicit in $V_{\textsf{min}} [f(r)]$, from evaluating $V_{\textsf{min}}$ with a $\lambda'$-corrected metric function, and those that are explicit.

Replacing the above expansions in (\ref{potentialz}) and expanding again in $\lambda'$, by separately considering the terms of order zero and first order in $\lambda'$ we obtain two separate differential equations, a homogeneous and a nonhomogeneous one:

\bea
\frac{d^2\psi_0}{dz^2} + \left(\omega^2 -V_0\right)\psi_0 &=& 0, \label{potentialz0} \\
\frac{d^2\psi_1}{dz^2} + \left(\omega^2 -V_0\right)\psi_1 &=& \xi, \label{potentialz1}
\eea
with the function $\xi$ given by
\bea
\xi &=& \xi_1+\xi_2+\xi_3, \label{xi} \\
\xi_1&=& -2 \delta f(r) \left(\frac{d^2 \psi_0}{dz^2}\right), \label{xi1} \\
\xi_2&=& - f(r) \left[\delta f(r) \right]'  \left(\frac{d\psi_0}{dz}\right), \label{xi2} \\
\xi_3&=&  V_1(r) \psi_0. \label{xi3}
\eea

Asymptotically as $|r| \to +\infty$, we can choose $ z(r) \sim x(r)$ \cite{Moura:2021nuh} and rewrite the boundary condition (\ref{130}) as
\begin{equation}
    \psi_\omega(z) \sim e^{i\omega z} +R(\omega)e^{-i\omega z}.
    \label{132}
\end{equation}

The strategy we use to compute the greybody factor consists in building a linear system of three algebraic equations where one of the independent variables is the reflection coefficient $R(\omega)$. After solving this system and consequently finding $R(\omega)$, we repeat the process to find $\widetilde{R}(\omega):= R(-\omega).$ Finally, we relate these coefficients with the greybody factor $\gamma(\omega)$, using equations (\ref{148}) and (\ref{149}).

\subsection{Tensorial gravitational perturbations}
\label{sec52}
\noindent

We now proceed with the calculation of the asymptotic greybody factor corresponding to tensorial gravitational perturbations, which are described by the potential (\ref{vt}).

\subsubsection{Computation of $R(\omega)$}
\noindent

In order to build the system we seek, we use the big and small contours defined in section \ref{gs}.

In an arbitrarily small neighborhood of the origin of the complex $r$-plane, we can then write the differential equation (\ref{potentialz0}) as
\begin{equation}
    \frac{d^2\psi_0}{dz^2} + \left(\omega^2 - \frac{j^2-1}{4z^2}\right)\psi_0 = 0 \label{psi0e}
\end{equation}
for $j =0$. Following the procedure of \cite{Motl:2003cd} we will consider the general solution, for arbitrary $j$, of the above differential equation, and at the end take the limit $j \rightarrow 0$. Such solution is given by
\begin{equation}
    \psi_0(z) = A_+ \sqrt{2\pi} \sqrt{\omega z} J_{\frac{j}{2}}(\omega z) + A_- \sqrt{2\pi} \sqrt{\omega z} J_{-\frac{j}{2}}(\omega z), \label{psi0}
\end{equation}
where $J_{\pm\frac{j}{2}}(\omega z)$ are Bessel functions of the first kind and $A_+,A_-$ arbitrary constants.

Close to the origin, $z$ can be approximated simply as
\begin{equation}
z \sim - \frac{1}{d-2}\frac{r^{d-2}}{R_H^{d-3}}. \label{z0}
\end{equation}
Moreover, for some fixed $A_\pm$, the particular solution to the nonhomogeneous differential equation (\ref{potentialz1}), obtained by the method of  variation of constants, can be decomposed into a sum of three terms, each one corresponding to a term $\xi_i$ in (\ref{xi}):
\bea
\psi_1(z) &=& \sum_{i=1}^3 \phi_i(z), \label{psi1} \\
\phi_i(z) &=& 2\pi \sqrt{\omega z}J_{-\frac{j}{2}}(\omega z) \int \sqrt{\omega z}J_{\frac{j}{2}}(\omega z)\frac{\xi_i(z)}{W}dz - 2\pi \sqrt{\omega z}J_{\frac{j}{2}}(\omega z) \int \sqrt{\omega z} J_{-\frac{j}{2}}(\omega z)\frac{\xi_i(z)}{W}dz, \label{fi}\\
W &=& -4\omega \sin\left(\frac{\pi j}{2}\right). \nonumber
\eea
In order to evaluate these functions, from (\ref{xi}), (\ref{fi}) we need to study the following class of indefinite integrals:
\begin{equation}
\mathcal{P}_{m n k}\left(x\right) = \int x^k J_m(x)J_n(x)dx, \label{pkmn}
\end{equation}
for $m,n \in \mathbb{R}$ and $k < 0$. These integrals are given in terms of generalized hypergeometric functions, having the following asymptotic behavior for $|x|\gg 1$, considering that $k < 0$:
\begin{equation}
    \mathcal{P}_{ m nk}\left(x\right) \sim \mathcal{H}(m,n,k) +\mathcal{O}(x^k),
    \label{expansion1}
\end{equation}
where
\begin{equation}
    \mathcal{H}(m,n,k):= \frac{\Gamma \left(\frac{1}{2}-\frac{k }{2}\right) \Gamma \left(-\frac{k }{2}\right) \Gamma \left(\frac{k }{2}+\frac{m }{2}+\frac{n }{2}+\frac{1}{2}\right)}{2 \sqrt{\pi } \Gamma \left(-\frac{k }{2}+\frac{m }{2}-\frac{n }{2}+\frac{1}{2}\right) \Gamma \left(-\frac{k }{2}+\frac{n }{2}-\frac{m }{2}+\frac{1}{2}\right) \Gamma \left(-\frac{k}{2}+\frac{m }{2}+\frac{n }{2}+\frac{1}{2}\right)}. \label{hkmn}
\end{equation}

Associated to each term $\xi_i$ in (\ref{xi}) corresponds an expansion in $z$ whose leading terms, expressed in terms of $z$ through (\ref{z0}), are given by
\bea
2 \delta f(r) &\sim& \Upsilon_1 \left(\frac{R_H}{z}\right)^{-\rho-2}, \\
f(r) \left[\delta f(r) \right]' &\sim& \Upsilon_2 z^{\rho+1} R_H^{-\rho-2}, \\
V_1(r) &\sim& \Upsilon_3 z^\rho R_H^{-\rho-2}, \label{v10}
\eea
with the definitions
\bea
\rho &\equiv& -2-\frac{d-1}{d-2}, \label{rho}\\
\Upsilon_1 &\equiv& (-1)^{\rho+2} (d-2)^{\rho+2} (d-4)(d-3), \label{u1}\\
\Upsilon_2 &\equiv& (-1)^{\rho+1} (d-2)^{\rho+1} \frac{(d-4)(d-3)(d-1)}{2},  \label{u2}\\
\Upsilon_3 &\equiv& \frac{1}{4} (-1)^\rho (d-2)^\rho (d-4) ((d-5) d (2 d-7)-22) \label{u3}.
\eea
We now introduce the definitions
\begin{equation}
    \Omega^\pm_i(d,j) = \pm 2^{i-4} \pi \Upsilon_i \csc\left(\frac{\pi j}{2}\right),\, i = 1, 2, 3
\end{equation}
and
\begin{equation}
\begin{split}
    \Theta_1^\pm(d,j) := \Omega^\pm_1 \left(A_\mp\mathcal{H}\left(\mp\frac{j}{2},\mp\frac{j}{2},\rho +1\right) + A_\pm \mathcal{H}\left(\mp\frac{j}{2},\pm\frac{j}{2},\rho +1\right)\right)\left(j^2-1\right)\\  -4\Omega^\pm_1 \left(A_\mp\mathcal{H}\left(\mp\frac{j}{2},\mp\frac{j}{2},\rho +3\right) + A_\pm \mathcal{H}\left(\mp\frac{j}{2},\pm\frac{j}{2},\rho +3\right)\right),
    \end{split}
\end{equation}
\begin{equation}
\begin{split}
   \Theta^\pm_2(d,j) := \Omega^\pm_2 \left(A_\mp\mathcal{H}\left(\mp\frac{j}{2},\mp\frac{j}{2},\rho+1\right) + A_\pm\mathcal{H}\left(\mp\frac{j}{2},\pm\frac{j}{2},\rho+1\right)\right) \\ \pm \Omega^\pm_2 \left[A_\mp\mathcal{H} \left(\mp\frac{j}{2},\mp\left(\frac{j}{2}+1\right),\rho+2\right) +A_\pm\mathcal{H}\left(\mp\frac{j}{2},\pm\left(\frac{j}{2}-1\right),\rho+2\right) \right.\\
   - \left.A_\mp\mathcal{H}\left(\mp\frac{j}{2},\mp\left(\frac{j}{2}-1\right),\rho+2\right) - A_\pm\mathcal{H}\left(\mp\frac{j}{2},\pm\left(\frac{j}{2}+1\right),\rho+2\right)\right],
   \end{split}
\end{equation}
\begin{equation}
\Theta_3^\pm(d,j) := \Omega^\pm_3 \left(A_\mp\mathcal{H}\left(\mp\frac{j}{2},\mp\frac{j}{2},\rho +1\right) + A_\pm \mathcal{H}\left(\mp\frac{j}{2},\pm\frac{j}{2},\rho +1\right)\right).
\end{equation}

Near point $D$ in figure \ref{fig2}, the analytical continuation of the solution $\psi_{\omega}(z)$ that we found near the origin (expressed through (\ref{psidef}), (\ref{psi0}) and (\ref{psi1})) is given by
\begin{equation}
\begin{split}
    \psi_{\omega}(z)  \sim
    \left(A_+e^{i\alpha_+}+A_-e^{i\alpha_-}\right)e^{-i\omega z}\left[1 + \lambda' \left(\frac{\Lambda_I^+e^{i\alpha_+} +\Lambda_I^-e^{i\alpha_-}}{A_+e^{i\alpha_+} +A_-e^{i\alpha_-}}\right)\right] \\ +
    \left(A_+e^{-i\alpha_+}+A_-e^{-i\alpha_-}\right)e^{i\omega z}\left[1 + \lambda' \left(\frac{\Lambda_I^+e^{-i\alpha_+} +\Lambda_I^-e^{-i\alpha_-}}{A_+e^{-i\alpha_+} +A_-e^{-i\alpha_-}}\right)\right],
    \label{173}
    \end{split}
\end{equation}
where $\alpha_\pm$ and $\Lambda^\pm_I$ are defined by
\bea
\alpha_ {\pm} &\coloneqq& \frac{\pi}{4}\left(1 \pm j\right), \label{105} \\
\Lambda_I^\pm(A_+,A_-) &=& \left(R_H \omega \right)^{\frac{d-1}{d-2}} \sum_{k=1}^3\Theta_k^\pm(d,j). \label{lambi}
\eea
We kept explicit the dependence of $\Lambda_I^\pm$ on the constants $A_+,A_-$ on the definition (\ref{lambi}) for reasons that will be clear later.

Imposing the boundary condition (\ref{132}) on the expression (\ref{173}) above yields the algebraic equations

\begin{equation}
   \left(A_++\lambda'\Lambda_I^+\right)e^{i\alpha_+} + \left(A_-+\lambda'\Lambda_I^-\right)e^{i\alpha_-} = R(\omega)
   \label{136}
\end{equation}
\begin{equation}
   \left(A_++\lambda'\Lambda_I^+\right)e^{-i\alpha_+} + \left(A_-+\lambda'\Lambda_I^-\right)e^{-i\alpha_-} = 1.
   \label{138}
\end{equation}

Now, we only need one equation to complete our system. This equation will encode exclusively functional information of the boundary condition (\ref{129}), turning a blind eye to $T(\omega)$. We build such equation using the monodromy theorem: indeed, we will equate two monodromies of $\psi_\omega$, associated with full clockwise loops around the big and small contours.

After following the big contour and returning to $D$, we have \cite{Moura:2021nuh}
\begin{equation}
    \psi_{\omega}(z) \sim \left(A_+e^{5i\alpha_+} + A_-e^{5i\alpha_-}\right) e^{-i\omega z} \left[1 + \lambda'\left(\frac{\Lambda_G^+e^{i 5\alpha_+} +\Lambda_G^-e^{i 5\alpha_-}}{A_+ e^{5i\alpha_+}+A_-e^{5i \alpha_-}}\right)\right].
    \label{119}
\end{equation}
The constants $\Lambda_G^\pm$ are defined as
\begin{equation}
    \Lambda_G^\pm(A_+,A_-):= \left(R_H \omega \right)^{\frac{d-1}{d-2}} \sum_{k=1}^3\Xi_k^\pm,
    \label{142}
\end{equation}
with
\bea
\Xi_1^+(d,j) &=& \Omega^\pm_1 e^{3\pi i (\rho + 2)} \left(j^2-1\right)\left[A_\mp e^{-3\pi ij} \mathcal{H}\left(\mp\frac{j}{2},\mp\frac{j}{2},\rho +1\right) + A_\pm \mathcal{H}\left(\mp\frac{j}{2}, \pm\frac{j}{2},\rho +1\right)\right] \label{133} \\
&-& 4\Omega^\pm_1  e^{3\pi i(\rho + 4)} \left[A_\mp e^{-3\pi i j} \mathcal{H}\left(\mp \frac{j}{2},\mp\frac{j}{2},\rho +3\right) + A_\pm \mathcal{H} \left(\mp\frac{j}{2},\pm\frac{j}{2},\rho +3\right)\right], \nonumber \\
\Xi^\pm_2(d,j) &=& \Omega^\pm_2 e^{3\pi i(\rho + 2)} \left[A_\mp e^{\mp 3\pi i j}\mathcal{H}\left(\mp\frac{j}{2},\mp\frac{j}{2},\rho+1\right) + A_\pm \mathcal{H}\left(\mp\frac{j}{2},\pm\frac{j}{2},\rho+1\right)\right]  \\
&+& \Omega^\pm_2 e^{3\pi i(\rho + 2)} \left[A_\mp e^{\mp 3\pi i j} \mathcal{H}\left(\mp \frac{j}{2},\mp \frac{j}{2}-1,\rho+2\right)+A_\pm \mathcal{H}\left(\mp \frac{j}{2},\pm \frac{j}{2}-1,\rho+2\right)\right] \nonumber \\
&-&\Omega_2^\pm e^{3\pi i (\rho+ 4)} \left[A_\mp e^{\mp 3\pi i j} \mathcal{H}\left(\mp \frac{j}{2},1\mp \frac{j}{2},\rho+2\right) + A_\pm \mathcal{H}\left(\mp\frac{j}{2},\pm \frac{j}{2}+1,\rho+2\right)\right], \nonumber \\
\Xi_3^\pm (d,j) &=& \Omega^\pm_3 e^{3\pi i(\rho + 2)} \left[A_\mp e^{-3\pi i j}\mathcal{H}\left(\mp \frac{j}{2},\mp \frac{j}{2},\rho +1\right) + A_\pm \mathcal{H}\left(\mp \frac{j}{2},\pm \frac{j}{2},\rho +1\right)\right].
\eea

Before computing the monodromy of $\psi$, we need to address one last detail: $z$ has a branch point in the real and fictitious horizons. Since the big contour encloses the real horizon, a full loop around it is bound to cross a branch cut somewhere. Thus, the expression (\ref{119}) above is written with respect to a variable $z$ defined on a branch of $z$ different from the branch where $z$ is defined for (\ref{173}). In order to relate these two variables, according to the redefinition $z \mapsto z + \Delta_z$, we need to consider the monodromy of $z$ associated with a full clockwise loop around $r = R_H$, given by \cite{Moura:2021nuh}
\begin{equation}
\Delta_z := -2\pi i \frac{R_H}{d-3}. \label{mm}
\end{equation}

Hence, the (multiplicative) monodromy of $\psi$, associated with a full clockwise loop around the big contour, is
\begin{equation}
    \mathcal{N}_1 := \left(\frac{A_+e^{5i\alpha_+} + A_-e^{5i\alpha_-}}{A_+e^{i\alpha_+} + A_-e^{i\alpha_-}}\right) e^{-i\omega \Delta_z}\left(1 + \lambda' \delta \mathcal{N}_1\right)
    \label{137}
\end{equation}
where we defined
\begin{equation}
    \delta \mathcal{N}_1 := \frac{\Lambda_G^+e^{5i\alpha_+} +\Lambda_G^-e^{5i\alpha_-}}{A_+ e^{5i\alpha_+}+ A_-e^{5i\alpha_-}} - \frac{\Lambda_I^+e^{i\alpha_+} +\Lambda_I^-e^{i\alpha_-}}{A_+e^{i\alpha_+} + A_-e^{i\alpha_-}}
\end{equation}

\begin{figure}[H]
\centering
\includegraphics[width=0.5\textwidth]{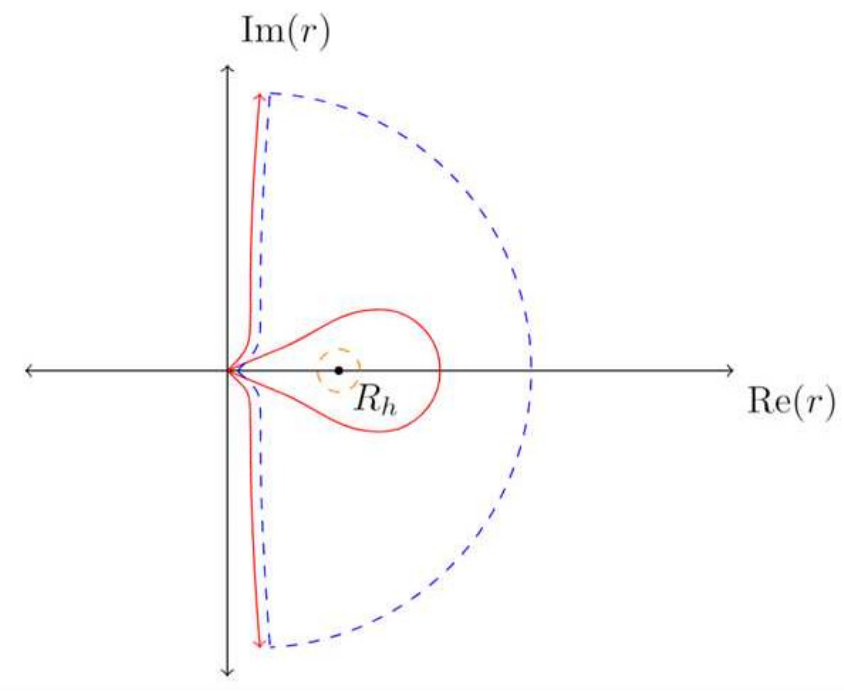}
\caption{Schematic depiction of the small and big contours as the orange and blue dashed lines respectively. The orange contour is to be interpreted as arbitrarily close to $R_H$. Red curves are Stokes lines.}
\label{fig4}
\end{figure}

The small contour is an arbitrarily small closed contour around the event horizon $R_H.$ It can be represented as the dashed orange contour in figure \ref{fig4}.

For the monodromy of $\psi_\omega$ associated with a full clockwise loop around the small contour we get the same result as in the calculation of asymptotic quasinormal modes. This is so, because the boundary condition (\ref{129}) has the same functional form as the equivalent boundary condition for quasinormal modes. Hence, we can obtain the monodromy of $\psi_\omega$, associated with a full clockwise loop around the small contour, simply exponentiating the monodromy of the tortoise coordinate $x$, which we designate by $\Delta_x$. This monodromy is obviously related to the monodromy $\Delta_z$ of $z$ given by (\ref{mm}) \cite{Moura:2021nuh}:
\begin{equation}
    \mathcal{N}_2 := e^{i\omega \Delta_x}, \,  \Delta_x= \Delta_z \left(1 + \frac{\lambda'}{2}(d-4)(d-1)\right).
    \label{174}
\end{equation}

Because the big and small contours are homotopic, the monodromy theorem yields the equation
\begin{equation}
    \mathcal{N}_1 = \mathcal{N}_2.
\end{equation}
Using (\ref{136}), (\ref{137}) and (\ref{174}), we can rewrite the equation above as
\begin{equation}
   R(\omega) = e^{-i\omega( \Delta_z+ \Delta_x)}\left[\left(A_+ +\lambda'\Lambda_G^+\right)e^{5i\alpha_+} +  \left(A_-+ \lambda'\Lambda_G^-\right)e^{5i\alpha_-}\right].
   \label{185}
\end{equation}

The equations (\ref{136}), (\ref{138}) and (\ref{185}) make up the linear system we seek. The independent variables are the reflection coefficient $R(\omega)$ together with the complex constants $A_\pm$.

In order to solve this system, we use standard perturbation theory. Thus, we start by considering the expansions
\begin{equation}
    R(\omega) = R_0(\omega) + \lambda'R_1(\omega)
    \label{144}
\end{equation}
\begin{equation}
    A_\pm = A_\pm^0+\lambda'A_\pm^1.
    \label{179}
\end{equation}
The monodromies $\Delta_z, \, \Delta_x$ can be written in terms of the black hole temperature $T_{\mathcal{H}}$ given by (\ref{temp}). Up to first order in $\lambda'$ we have
\begin{equation}
    \Delta_x+ \Delta_z = -\frac{i}{T_{\mathcal{H}}}\left(1-\lambda'\frac{(d-1)(d-4)}{4}\right).
    \label{181}
\end{equation}
Furthermore, using (\ref{181}) allow us to write the Taylor expansion
\begin{equation}
    e^{-i\omega(\Delta_z+\Delta_x)} \sim e^{-\frac{\omega}{T_{\mathcal{H}}}}\left(1+ \lambda'\frac{\omega}{T_{\mathcal{H}}}\frac{(d-4)(d-1)}{4}\right)
\end{equation}
again up to first order in $\lambda'$. Replacing the expansions above in the equations of our system and solving them perturbatively in powers of $\lambda'$ yields two distinct linear systems of algebraic equations. The first one, of zeroth order in $\lambda'$, is
\begin{equation}
    \begin{dcases}
    A_+^0e^{5i\alpha_+} + A_-^0e^{5i\alpha_-} =  R_0(\omega) e^{\frac{\omega}{T_{\mathcal{H}}}} \\
    A_+^0e^{i\alpha_+} + A_-^0e^{i\alpha_-} = R_0(\omega) \\
    A_+^0e^{-i\alpha_+} + A_-^0e^{-i\alpha_-} = 1
    \end{dcases}
\end{equation}
whose solution is
\begin{equation}
    R_0(\omega) =  \frac{2i\cos\left(\frac{\pi j}{2}\right)}{e^{\frac{\omega}{T_{\mathcal{H}}}}+2\cos(\pi j)+1},
    \label{145}
\end{equation}
\begin{equation}
    A_+^0 = \frac{e^{i\alpha_-} - R_0(\omega)e^{-i\alpha_-}}{2 i \sin\left(\alpha_--\alpha_+\right)},
    \label{177}
\end{equation}
\begin{equation}
    A_-^0= \frac{R_0(\omega)e^{-i\alpha_+}- e^{i\alpha_+}}{2i\sin\left(\alpha_--\alpha_+\right)}.
    \label{183}
\end{equation}
We now define $\zeta_\pm,\sigma_\pm$ as the constants resulting from replacing $A_+$ and $A_-$ in the definitions (\ref{lambi}) of $\Lambda_I^\pm$ (respectively (\ref{142}) of $\Lambda_G^\pm$) by the results (\ref{177}) and (\ref{183}) for $A_+^0$ and $A_-^0$. Symbolically,
\bea
\zeta_\pm := \Lambda_I^\pm(A_+^0,A_-^0), \\
\sigma_\pm := \Lambda_G^\pm(A_+^0,A_-^0).
\eea

The second linear system of algebraic equations, of first order in $\lambda'$, is
\begin{equation}
    \begin{dcases}  \left(A_+^1 + \sigma_+\right)e^{5i\alpha_+} + \left(A_-^1+ \sigma_-\right)e^{5i\alpha_-} =   e^{\frac{\omega}{T_{\mathcal{H}}}} \left(R_1(\omega) - \frac{(d-4)(d-1)}{4} \omega T_{\mathcal{H}}^{-1} R_0(\omega)\right)\\
    \left(A_+^1+ \zeta_+\right)e^{i\alpha_+} + (A_-^1 + \zeta_-)e^{i\alpha_-} = R_1(\omega) \\
    \left(A_+^1+ \zeta_+\right)e^{-i\alpha_+} + (A_-^1 + \zeta_-)e^{-i\alpha_-} = 0.
    \end{dcases}
    \label{140}
\end{equation}
Moreover, defining the constants
\begin{equation}
    \Sigma_1 := -\left(\zeta_+e^{i\alpha_+} +\zeta_-e^{i\alpha_-} \right),
\end{equation}
\begin{equation}
    \Sigma_2 := -\left(\zeta_+e^{-i\alpha_+} +\zeta_-e^{-i\alpha_-} \right),
\end{equation}
\begin{equation}
    \Sigma_3 :=
     -\frac{(d-4)(d-1)}{4}\omega T_{\mathcal{H}}^{-1}e^{\frac{\omega}{T_{\mathcal{H}}}} R_0(\omega) -\left(\sigma_+e^{5i\alpha_+} +\sigma_-e^{5i\alpha_-}\right),
\end{equation}
we can rewrite the system (\ref{140}) in the simple form
\begin{equation}
    \begin{dcases}
   A_+^1e^{5i\alpha_+} + A_-^1e^{5i\alpha_-}  - R_1(\omega)e^{\frac{\omega}{T_{\mathcal{H}}}} = \Sigma_3\\
    A_+^1e^{i\alpha_+} + A_-^1e^{i\alpha_-} - R_1(\omega) = \Sigma_1 \\  A_+^1e^{-i\alpha_+} + A_-^1e^{-i\alpha_-}  = \Sigma_2.
    \end{dcases}
\end{equation}
After solving this system for $R_1(\omega),$ we can write (\ref{144}) in the form
\begin{equation}
    R(\omega) = R_0(\omega)\left(1 + \lambda'\delta R_g(\omega)\right) \label{drg}
\end{equation}
with $R_0(\omega)$ given by (\ref{145}) and
\begin{equation}
    \delta R_g(\omega) := -\frac{2\Sigma_1\cos(\pi j)-2i\Sigma_2\cos\left(\frac{\pi j}{2}\right) + \Sigma_1 + \Sigma_3}{R_0(\omega)\left(e^{\frac{\omega}{T_{\mathcal{H}}}}+2\cos(\pi j)+1\right)}.
\end{equation}
Taking the limit $j\to 0$ and performing a large amount of algebraic manipulation yields
\begin{equation}
    R_0(\omega) = \frac{2i}{e^{\frac{\omega}{T_{\mathcal{H}}}}+3},
\end{equation}
\begin{equation}
    \delta R_g(\omega) =\frac{1}{e^{\frac{\omega}{T_{\mathcal{H}}}}+3} \left[\frac{(d-4)(d-1)}{4}\frac{\omega}{T_{\mathcal{H}}}e^{\frac{\omega}{T_\mathcal{H}}}+ \left(\frac{\omega}{T_{\mathcal{H}}}\right)^{\frac{d-1}{d-2}}\left(\frac{d-3}{4\pi}\right)^{\frac{d-1}{d-2}}\Upsilon_g\right]
\end{equation}
with
\begin{equation}
    \Upsilon_g \coloneqq e^{-\frac{3 \pi i}{2}\left(1+ \frac{1}{d-2}\right)} \frac{\pi ^2 (2d-4)^{-\frac{1}{d-2}} (d-4) (d-1)^3 ((d-5) d+2)  \Gamma \left(\frac{1}{d-2}\right) \left(e^{-\frac{i \pi}{d-2}} \left(e^{\frac{\omega}{T_{\mathcal{H}}}}-1\right)+4\right)}{16 (d-2)^5 \Gamma \left(\frac{1}{2}\left(3 + \frac{1}{d-2}\right)\right)^4}.
\end{equation}

\subsubsection{Computation of $\widetilde{R}(\omega)$}
\noindent

In order to compute $\widetilde{R}(\omega)$, we take the same procedure used in the previous section, but considering the solution $\psi_{-\omega}$ of (\ref{potential0}). This solution has an expansion in $\lambda'$ analogous to (\ref{psidef}), which we consider together with (\ref{v1def}):
\be
\psi_{-\omega} = \widetilde{\psi}_0 + \lambda' \widetilde{\psi}_1. \label{psitildef}
\ee

The boundary condition (\ref{130til}) can be written with respect to $z$ as
\begin{equation}
    \psi_{-\omega}(z) \sim e^{-i\omega z} +\widetilde{R}(\omega)e^{i\omega z},
    \label{141}
\end{equation}
for $r \to + \infty$, on the same Stokes line.

Like (\ref{psi0}), we still can write the asymptotic expansion near the origin
\begin{equation}
    \widetilde{\psi}_0 (z) \sim \widetilde{A}_+\sqrt{2\pi}\sqrt{\omega z} J_{\frac{j}{2}}(\omega z) + \widetilde{A}_-\sqrt{2\pi}\sqrt{\omega z} J_{-\frac{j}{2}}(\omega z),
\end{equation}
for some arbitrary constants $\widetilde{A}_\pm \in \mathbb{C}$. In this case, analogously to (\ref{173}) we can write
\begin{equation}
\begin{split}
    \psi_{-\omega}(z) \sim
    \left(\widetilde{A}_+e^{i\alpha_+}+\widetilde{A}_-e^{i\alpha_-}\right)e^{-i\omega z}\left[1 + \lambda' \left(\frac{\widetilde{\Lambda}_I^+e^{i\alpha_+} +\widetilde{\Lambda}_I^-e^{i\alpha_-}}{\widetilde{A}_+e^{i\alpha_+} +\widetilde{A}_-e^{i\alpha_-}}\right)\right] \\ +
    \left(\widetilde{A}_+e^{-i\alpha_+}+\widetilde{A}_-e^{-i\alpha_-}\right)e^{i\omega z}\left[1 + \lambda' \left(\frac{\widetilde{\Lambda}_I^+e^{-i\alpha_+} +\widetilde{\Lambda}_I^-e^{-i\alpha_-}}{\widetilde{A}_+e^{-i\alpha_+} +\widetilde{A}_-e^{-i\alpha_-}}\right)\right]
    \end{split}
\end{equation}
near $D$, where $\widetilde{\Lambda}_I^\pm=\Lambda_I^\pm(\widetilde{A}_+,\widetilde{A}_-)$ are the constants resulting from switching $A_\pm$ for $\widetilde{A}_\pm$ in the definitions (\ref{lambi}). Imposing the boundary condition (\ref{141}) on the expression above yields the algebraic equations
\bea
\left(\widetilde{A}_++\lambda'\widetilde{\Lambda}_I^+\right)e^{i\alpha_+} + \left(\widetilde{A}_-+\lambda'\widetilde{\Lambda}_I^-\right)e^{i\alpha_-} = 1, \label{175} \\
\left(\widetilde{A}_++\lambda'\widetilde{\Lambda}_I^+\right)e^{-i\alpha_+} + \left(\widetilde{A}_-+\lambda'\widetilde{\Lambda}_I^-\right)e^{-i\alpha_-} = \widetilde{R}(\omega). \label{178}
\eea

Analogously to (\ref{137}), the monodromy of $\psi_{-\omega}$, associated with a full clock wise loop around the big contour, is
\begin{equation}
    \widetilde{\mathcal{N}}_1 := \left(\frac{\widetilde{A}_+e^{5i\alpha_+} + \widetilde{A}_-e^{5i\alpha_-}}{\widetilde{A}_+e^{i\alpha_+} + \widetilde{A}_-e^{i\alpha_-}}\right) e^{-i\omega \Delta_z}\left(1 + \lambda' \delta \widetilde{\mathcal{N}}_1\right)
    \label{143}
\end{equation}
where we defined
\begin{equation}
    \delta \widetilde{\mathcal{N}}_1 := \frac{\widetilde{\Lambda}_G^+e^{i5\alpha_+} +\widetilde{\Lambda}_G^-e^{i5\alpha_-}}{\widetilde{A}_+ e^{5i\alpha_+}+ \widetilde{A}_-e^{5i\alpha_-}} - \frac{\widetilde{\Lambda}_I^+e^{i\alpha_+} +\widetilde{\Lambda}_I^-e^{i\alpha_-}}{\widetilde{A}_+e^{i\alpha_+} + \widetilde{A}_+e^{i\alpha_-}}.
\end{equation}
In the expression above, $\widetilde{\Lambda}^\pm_G=\Lambda_G^\pm(\widetilde{A}_+,\widetilde{A}_-)$ denotes the constants resulting from switching $A_\pm$ for $\widetilde{A}_\pm$ in the definition (\ref{142}).

From the boundary condition (\ref{129til}) for $r \to R_H$ we get, analogously to (\ref{174}),
\begin{equation}
    \widetilde{\mathcal{N}}_2 := e^{-i\omega \Delta_ x}
    \label{146}
\end{equation}
as the multiplicative monodromy of $\psi_{-\omega}$ associated with a full clockwise loop around the small contour.

Using the monodromy theorem yields the equation
\begin{equation}
    \widetilde{\mathcal{N}}_1 =  \widetilde{\mathcal{N}}_2.
\end{equation}
Using (\ref{175}), (\ref{143}) and (\ref{146}), we can rewrite the equation above as
\begin{equation}
    1 = e^{-i\omega(\Delta_z-\Delta_x)}\left[\left(\widetilde{A}_+ +\lambda'\widetilde{\Lambda}_G^+\right)e^{5i\alpha_+} +  \left(\widetilde{A}_-+ \lambda'\widetilde{\Lambda}_G^-\right)e^{5i\alpha_-}\right].
    \label{147}
\end{equation}
Furthermore, from (\ref{temp}), (\ref{mm}) and (\ref{174}) we can write the Taylor expansion
\begin{equation}
    e^{-i\omega(\Delta_z-\Delta_x)} \sim 1 + \lambda' \frac{(d-4)(d-1)}{4} \frac{\omega}{T_{\mathcal{H}}}
\end{equation}
up to first order in $\lambda'$.

The equations (\ref{175}), (\ref{178}) and (\ref{147}) make up the linear system of algebraic equations we seek. Once again, we address this system using standard perturbation theory. Thus, we start by considering the expansions
\begin{equation}
    \widetilde{R}(\omega) = \widetilde{R}_0(\omega) + \lambda'\widetilde{R}_1(\omega).
    \label{184}
\end{equation}
\begin{equation}
    \widetilde{A}_\pm = \widetilde{A}^0_\pm + \lambda' \widetilde{A}^1_\pm
\end{equation}
Plugging the expansions above in the equations of our system and solving them perturbatively in powers of $\lambda'$ yields two distinct linear systems of algebraic equations. The first one, of zeroth order in $\lambda'$, is
\begin{equation}
    \begin{dcases}  \widetilde{A}_+^0e^{5i\alpha_+} + \widetilde{A}_-^0e^{5i\alpha_-} =  1,\\
    \widetilde{A}_+^0e^{i\alpha_+} + \widetilde{A}_-^0e^{i\alpha_-} = 1,\\
    \widetilde{A}_+^0e^{-i\alpha_+} + \widetilde{A}_-^0e^{-i\alpha_-} = \widetilde{R}(\omega),
    \end{dcases}
\end{equation}
whose solution is
\begin{equation}
    \widetilde{R}_0(\omega) = -2i\cos\left(\frac{\pi j}{2}\right), \label{r0til}
\end{equation}
\begin{equation}
    \widetilde{A}_+^0 = \frac{-e^{-i\alpha_-} + \widetilde{R}_0(\omega)e^{i\alpha_-}}{2 i \sin\left(\alpha_--\alpha_+\right)},
    \label{172}
\end{equation}
\begin{equation}
    \widetilde{A}_-^0= \frac{-\widetilde{R}_0(\omega)e^{i\alpha_+}+ e^{-i\alpha_+}}{2i\sin\left(\alpha_--\alpha_+\right)}.
    \label{180}
\end{equation}
The second system, of first order in $\lambda'$, is
\begin{equation}
    \begin{dcases}  \left(\widetilde{A}_+^1 + \widetilde{\sigma}_+\right)e^{5i\alpha_+} + \left(\widetilde{A}_-^1+ \widetilde{\sigma}_-\right)e^{5i\alpha_-} =    - \frac{(d-4)(d-1)}{4} \omega T_{\mathcal{H}}^{-1} \\
    \left(\widetilde{A}_+^1+ \widetilde{\zeta}_+\right)e^{i\alpha_+} + \left(\widetilde{A}_-^1 + \widetilde{\zeta}_-\right)e^{i\alpha_-} = 0 \\
    \left(\widetilde{A}_+^1+ \widetilde{\zeta}_+\right)e^{-i\alpha_+} + \left(\widetilde{A}_-^1 + \widetilde{\zeta}_-\right)e^{-i\alpha_-} = \widetilde{R}_1(\omega)
    \label{176}
    \end{dcases}
\end{equation}
where we defined $\widetilde{\zeta}_\pm, \widetilde{\sigma}_\pm$, analogously to $\zeta_\pm,\sigma_\pm$, as the constants resulting from switching $\widetilde{A}_+$ and $\widetilde{A}_-$ in the definitions of $\widetilde{\Lambda}_I^\pm, \widetilde{\Lambda}_G^\pm$ by the results (\ref{172}) and (\ref{180}) for $\widetilde{A}_+^0$ and $\widetilde{A}_-^0$ respectively. Symbolically,
\bea
\widetilde{\zeta}_\pm := \widetilde{\Lambda}_I^\pm(\widetilde{A}_+^0,\widetilde{A}_-^0), \\
\widetilde{\sigma}_\pm := \widetilde{\Lambda}_G^\pm(\widetilde{A}_+^0,\widetilde{A}_-^0).
\eea

Moreover, defining the constants
\begin{equation}
    \widetilde{\Sigma}_1 := -\left(\widetilde{\zeta}_+e^{i\alpha_+} +\widetilde{\zeta}_-e^{i\alpha_-} \right),
\end{equation}
\begin{equation}
     \widetilde{\Sigma}_2 := -\left(\widetilde{\zeta}_+e^{-i\alpha_+} +\widetilde{\zeta}_-e^{-i\alpha_-} \right),
\end{equation}
\begin{equation}
     \widetilde{\Sigma}_3 :=
     -\frac{(d-4)(d-1)}{4}\omega T_{\mathcal{H}}^{-1} -\widetilde{\sigma}_+e^{5i\alpha_+} -\widetilde{\sigma}_-e^{5i\alpha_-}
\end{equation}
allow us to rewrite (\ref{176}) as the simple system
\begin{equation}
    \begin{dcases}
   \widetilde{A}_+^1e^{5i\alpha_+} + \widetilde{A}_-^1e^{5i\alpha_-}   = \widetilde{\Sigma}_3\\
    \widetilde{A}_+^1e^{i\alpha_+} + \widetilde{A}_-^1e^{i\alpha_-}  = \widetilde{\Sigma}_1 \\  \widetilde{A}_+^1e^{-i\alpha_+} + \widetilde{A}_-^1e^{-i\alpha_-} - \widetilde{R}_1(\omega)  = \widetilde{\Sigma}_2.
    \end{dcases}
\end{equation}
Solving this system for $\widetilde{R}_1(\omega)$ yields
\begin{equation}
   \widetilde{R}_1(\omega) =  -\frac{i}{2}\sec\left(\frac{\pi j}{2}\right)\left(2\widetilde{\Sigma}_1\cos\left(\pi j\right) - 2i\widetilde{\Sigma}_2\cos\left(\frac{\pi j}{2}\right) + \widetilde{\Sigma}_1 + \widetilde{\Sigma}_3\right).
\end{equation}

Using (\ref{184}) and the equation above, we can write
\begin{equation}
    \widetilde{R}(\omega) = \widetilde{R}_0(\omega)\left(1 + \lambda' \delta \widetilde{R}_g(\omega)\right) \label{drgtil}
\end{equation}
with $\widetilde{R}_0(\omega)$ given by (\ref{r0til}), and where we defined
\begin{equation}
    \delta \widetilde{R}_g(\omega) :=  -\frac{i}{2\widetilde{R}_0(\omega)}\sec\left(\frac{\pi j}{2}\right)\left(2\widetilde{\Sigma}_1\cos\left(\pi j\right) - 2i\widetilde{\Sigma}_2\cos\left(\frac{\pi j}{2}\right) + \widetilde{\Sigma}_1 + \widetilde{\Sigma}_3\right).
\end{equation}
Taking the limit $j \to 0$ and performing a large amount of algebraic manipulation yields
\begin{equation}
    \widetilde{R}_0(\omega) = -2i
\end{equation}
\begin{equation}
    \delta\widetilde{R}_g(\omega) = -\frac{(d-1)(d-4)}{16} \frac{\omega}{T_{\mathcal{H}}} + \left(\frac{\omega}{T_{\mathcal{H}}}\right)^{\frac{d-1}{d-2}}\left(\frac{d-3}{4\pi}\right)^{\frac{d-1}{d-2}}\widetilde{\Upsilon}_g
\end{equation}
where we defined
\begin{equation}
   \Tilde{\Upsilon}_g \coloneqq -e^{-\frac{3\pi i}{2}\left(1 + \frac{1}{d-2}\right)} \frac{\pi ^2 2^{-\frac{1}{d-2}} (d-2)^{-\left(\frac{1}{d-2}+1\right)} (d-4) ((d-5) d+2)  \Gamma \left(\frac{1}{d-2}\right)}{(d-1) \Gamma \left(\frac{1}{2} \left(1+\frac{1}{d-2}\right)\right)^4}.
\end{equation}

\subsubsection{Computation of $\gamma(\omega)$}
\noindent

From the definition (\ref{148}) we can finally write the greybody factor as
\begin{equation}
    \gamma(\omega) = 1-R(\omega)\widetilde{R}(\omega) = \gamma_0(\omega)\left(1 + \lambda' \delta \gamma_g(\omega)\right)
    \label{215}
\end{equation}
with
\begin{equation}
    \gamma_0(\omega) :=  \frac{e^{\frac{\omega}{T_{\mathcal{H}}}}-1}{e^{\frac{\omega}{T_{\mathcal{H}}}}+3} \label{g0}
\end{equation}
\begin{equation}
    \delta\gamma_g(\omega) := \frac{4}{1-e^{\frac{\omega}{T_{\mathcal{H}}}}}\left(\delta\widetilde{R}_g(\omega) + \delta R_g(\omega)\right).
\end{equation}
After some algebraic manipulation, we can write
\begin{equation}
    \delta \gamma_g (\omega) =-\frac{4}{e^{\frac{\omega}{T_{\mathcal{H}}}}+3}\left[\frac{3}{16}(d-4)(d-1)\frac{\omega}{T_\mathcal{H}} +\left(\frac{\omega}{T_{\mathcal{H}}}\right)^{\frac{d-1}{d-2}} e^{-\frac{2\pi i}{d-2}} \varrho_g\right]
\end{equation}
with
\begin{equation}
    \varrho_g \coloneqq \pi^{\frac{d-4}{2(d-2)}} \left(\frac{d-3}{4(d-2)}\right)^{\frac{d-1}{d-2}} \frac{(d-4) ((d-5) d+2)}{(d-1)} \frac{\Gamma \left(\frac{1}{2(d-2)}\right)}{\Gamma \left(\frac{1}{2} + \frac{1}{2(d-2)}\right)^3}
    \sin\left(\frac{\pi}{2(d-2)}\right).
\end{equation}

It is interesting to study the magnitudes of the different contributions to the $\lambda'$ correction $\delta \gamma_g (\omega)$. For that purpose, we have evaluated $\varrho_g$ numerically for the relevant values of $d$, from $d=5$ (the smallest value of $d$ for which tensorial gravitational perturbations exist) to $d=10$ (where superstring theories are defined). This factor grows monotonically with $d$, varying from approximately 0.062 (corresponding to $d=5$) to approximately 7.349 (corresponding to $d=10$). Just for comparison, $\frac{3}{16}(d-1)(d-4)$ varies between 0.75 and $10.125$ in the same range.

\subsection{Test scalar fields}
\noindent

The computation of the asymptotic greybody factor of scalar test fields is completely analogous to the one corresponding to tensorial gravitational perturbations we have just seen, the only difference being the replacement of (\ref{vt}) by the potential (\ref{v0}). This corresponds to replacing $\Upsilon_3$ in (\ref{u3}) by the new value
\begin{equation}
\Upsilon_3 \mapsto \Upsilon_3 = \frac{1}{4} (-1)^\rho (d-2)^{\rho+1} (d-4)(d-3)(2d-3). \label{191}
\end{equation}
After repeating all the procedure and calculations of section \ref{sec52} with this replacement, we are led to new values of the reflection coefficients. Instead of (\ref{drg}) we now have
\begin{equation}
R(\omega) = R_0(\omega)\left(1+\lambda' \delta R_s (\omega)\right),
\end{equation}
with
\bea
\delta R_s(\omega) &\coloneqq& \frac{1}{e^{\frac{\omega}{T_{\mathcal{H}}}}+3} \left[\frac{(d-4)(d-1)}{4}\frac{\omega}{T_{\mathcal{H}}}e^{\frac{\omega}{T_\mathcal{H}}}+ \left(\frac{\omega}{T_{\mathcal{H}}}\right)^{\frac{d-1}{d-2}}\left(\frac{d-3}{4\pi}\right)^{\frac{d-1}{d-2}}\Upsilon_s\right],\\
\Upsilon_s &\coloneqq& e^{-\frac{3\pi i}{2}\left(1 + \frac{1}{d-2}\right)}\frac{\pi ^2 (2d-4)^{-\frac{1}{d-2}} (d-4) (d-3)   \Gamma \left(\frac{1}{d-2}\right) \left(e^{-\frac{i\pi}{d-2}} \left(e^{\frac{\omega}{T_{\mathcal{H}}}}-1\right)+4\right)}{(d-1) \Gamma \left(\frac{1}{2} + \frac{1}{2(d-2)}\right)^4}.
\eea
Additionally, instead of (\ref{drgtil}) we get
\begin{equation}
\widetilde{R}(\omega) = \widetilde{R}_0(\omega)\left(1 + \lambda' \delta \widetilde{R}_s(\omega)\right)
\end{equation}
with
\bea
\delta\widetilde{R}_s(\omega) &\coloneqq& -\frac{(d-1)(d-4)}{16} \frac{\omega}{T_{\mathcal{H}}} + \left(\frac{\omega}{T_{\mathcal{H}}}\right)^{\frac{d-1}{d-2}}\left(\frac{d-3}{4\pi}\right)^{\frac{d-1}{d-2}}\widetilde{\Upsilon}_s, \\
\widetilde{\Upsilon}_s &\coloneqq& - e^{-\frac{3\pi i}{2}\left(1 + \frac{1}{d-2}\right)}\frac{2 ^{-\frac{1}{d-2}}\pi ^2 (d-2)^{-\left(\frac{1}{d-2} + 4\right)} (d-4) (d-3) (d-1)^3   \Gamma \left(\frac{1}{d-2}\right)}{16  \Gamma \left(\frac{1}{2}\left(3 + \frac{1}{d-2}\right)\right)^4}.
\eea
Finally, we get for the greybody factor
\begin{equation}
    \gamma(\omega) =1- R(\omega)\widetilde{R}(\omega)= \gamma_0(\omega)\left(1+ \lambda' \delta \gamma_s(\omega)\right)
    \label{216}
\end{equation}
with $\gamma_0(\omega)$ given by (\ref{g0}) and
\bea
\delta \gamma_s (\omega) &\coloneqq& -\frac{4}{e^{\frac{\omega}{T_{\mathcal{H}}}}+3}\left(\frac{3}{16}(d-4)(d-1)\frac{\omega}{T_{\mathcal{H}}} + \left(\frac{\omega}{T_{\mathcal{H}}}\right)^{\frac{d-1}{d-2}} e^{-\frac{2\pi i}{d-2}} \varrho_s \right),\\
\varrho_s &\coloneqq& \left(\frac{d-3}{4\pi}\right)^{\frac{d-1}{d-3}} \frac{\pi^{\frac{3}{2}} (d-4) (d-3) \Gamma \left(\frac{1}{2 (d-2)}\right)}{(d-2)^{\frac{1}{d-2}} (d-1) \Gamma \left(\frac{1}{2}+\frac{1}{2(d-2)}\right)^3}\sin\left(\frac{\pi}{2(d-2)}\right).
\eea

Like we did for $\delta \gamma_g (\omega)$, we study the magnitudes of the different contributions to the $\lambda'$ correction $\delta \gamma_s (\omega)$. For that purpose, we have evaluated $\varrho_s$ numerically for the relevant values of $d$. This factor grows monotonically with $d$, varying from approximately 0.055 (corresponding to $d=5$) to approximately 7.204 (corresponding to $d=10$). These values are of the same order of magnitude of the ones obtained for tensorial gravitational perturbations.

\section{Conclusions}
\noindent

In this article, we have computed analytically the greybody factors relative to the emission of gravitons (corresponding to tensorial gravitational perturbations of the metric) and scalar test fields for the simplest case of a $d-$dimensional spherically symmetric black hole solution with leading string $\a$ corrections obtained by Callan, Myers and Perry \cite{cmp89}. We have considered complex emission frequencies, and we have taken high frequency limits: both the eikonal limit - where the real part of the frequency of the scattered wave is much larger than the imaginary part -, and the asymptotic highly damped case - where the imaginary part of the frequency is much larger than the real part. We have reobtained the classical part of these factors (corresponding to Einstein gravity), and we have computed the leading string $\a$ corrections. In both cases, and in both limits we observed that the corrections are strongly dependent on the spacetime dimension $d$. Moreover, they remain numerically small for every relevant value of $d$, not growing arbitrarily - and considering that they are multiplied by the naturally small inverse string tension $\a$.

Naturally a full study of greybody factors for black holes with higher derivative corrections should address the complete range of frequencies, and not only the limiting cases like the high frequency limits we have considered in this work and the low frequency limits that have been considered namely in \cite{Grain:2005my,Moura:2006pz,Moura:2011rr,Zhang:2017yfu}. Nonetheless, the knowledge of the analytical results corresponding to these limiting cases is important in order to be confronted with more complete numerical studies. Other approaches to the calculations of greybody factors, using symmetries of the master differential equation have been used for asymptotically flat black holes in $d=4$ \cite{Lenzi:2022wjv,Lenzi:2023inn}. In \cite{Bonelli:2021uvf} greybody factors of Kerr black holes have been obtained by computing the exact connection coefficients of the radial and angular parts of the Teukolsky equation. It would certainly be interesting to extend such studies to higher dimensions, and to the string corrected black holes we have considered.

\paragraph{Acknowledgements}
\noindent
This work has been supported by Funda\c c\~ao para a Ci\^encia e a Tecnologia under contracts IT (UIDB/50008/2020 and UIDP/50008/2020), CAMGSD/IST-ID (UIDB/04459/2020 and UIDP/04459/2020) and projects 2022.08368.PTDC and 2024.04456.CERN. Jo\~ao Rodrigues is supported by Funda\c c\~ao para a Ci\^encia e a Tecnologia through the doctoral fellowship UI/BD/151499/2021.

\end{document}